\documentclass[namedreferences]{SolarPhysics}
\usepackage[optionalrh]{spr-sola-addons} 
\usepackage{graphicx}        
\usepackage{color}           
\usepackage{url}             

\begin{document}

\begin{article}

\begin{opening}

\title{{\bf Multiwavelength Study of M8.9/3B Solar Flare from AR NOAA 10960}}

\author{Pankaj~\surname{Kumar}$^{1}$\sep
        A.K.~\surname{Srivastava}$^{1}$\sep
	B.~\surname{Filippov}$^{2}$\sep
        Wahab~\surname{Uddin}$^{1}$    
       }
\runningauthor{P.~Kumar {\it et al.}}
\runningtitle{Solar Flare from AR NOAA 10960}
\institute{$^{1}$ Aryabhatta Research Institute of Observational Sciences (ARIES), Manora Peak, Nainital-263129, India\\
                     email: \url{pkumar@aries.res.in}}
\institute{$^{2}$ Pushkov Institute of Terrestrial Magnetism, Ionosphere and Radio Wave Propagation, Russian Academy of
       Sciences, Troitsk Moscow Region 142190, Russia}

\begin{abstract}
We present a multiwavelength analysis of a long duration white-light solar flare (M8.9/3B) event that occurred on 04 June 2007 from AR NOAA 10960. The flare was observed by several spaceborne instruments, namely SOHO/MDI, HINODE/SOT, TRACE and STEREO/SECCHI. The flare was initiated near a small, positive-polarity, satellite sunspot at the centre of the active region, surrounded by opposite-polarity field regions. MDI images of the active region show considerable amount of changes in the small positive-polarity sunspot of $\delta$ configuration during the flare event. SOT/G-band (4305 \AA) images of the sunspot also suggest the rapid evolution of this positive-polarity sunspot with highly twisted penumbral filaments before the flare event, which were oriented in a counterclockwise direction. It shows the change in orientation, and also remarkable disapperance of twisted penumbral filaments ($\approx$35\,--\,40$\%$) and enhancement in umbral area ($\approx$45\,--50\,$\%$) during the decay phase of the flare event. TRACE and SECCHI observations reveal the successive activation of two helical-twisted structures associated with this sunspot, and the corresponding brightening in the chromosphere as observed by the time-sequence images of SOT/Ca\,{\sc ii} H line (3968 \AA). The secondary, helical-twisted structure is found to be associated with the M8.9 flare event. The brightening starts \,six--seven\, minutes prior to the flare maximum with the appearance of secondary, helical-twisted structure. The flare intensity maximizes as the secondary, helical-twisted structure moves away from the active region. This twisted flux tube, associated with the flare triggering, is found to be failed in eruption. The location of the flare activity is found to coincide with the activation site of the helical twisted structures. We conclude that the activation of successive helical twists (especially the second one) in the magnetic flux tubes/ropes plays a crucial role in the energy build-up process and triggering of the M-class solar flare without a coronal mass ejection (CME).
\end{abstract}
\end{opening}

\section{Introduction}
 Solar flares are the sudden explosions in the solar atmosphere during which large amounts of the magnetic energy, stored in the twisted and sheared magnetic fields, is released by the process of magnetic reconnection in the form of thermal energy and particle acceleration. The flares associated with the CME eruptions are known as ``eruptive flares'', while the flares without association of CMEs are known as ``confined flares''. The emerging magnetic flux,  rapid  motion/rotation of sunspots and interaction of filaments can destablize the magnetic field, and trigger the solar eruptive phenomena, {\it e.g.}, flares, CMEs {\it etc.} \cite{min2009,kumar2010}.  The S-shaped {\it or} inverted S-shaped sigmoids indicate the twisted field lines in the solar atmosphere.  The solar eruptions usually take place due to the increase of the twist in the magnetic field configurations of an active regions \cite {canfield1999}. Recently, the MHD models of magnetic flux tubes show that the twist [$\phi$] of \,2.5--\,3.5$\pi$ is sufficient for solar eruptions, and the stable equilibrium of the magneto-fluid breaks when the total twist in the associated flux tubes crosses this critical value \cite{fan2003,kliem2004,torok2004}.


\begin{figure}
\centering 
\includegraphics[width=8cm]{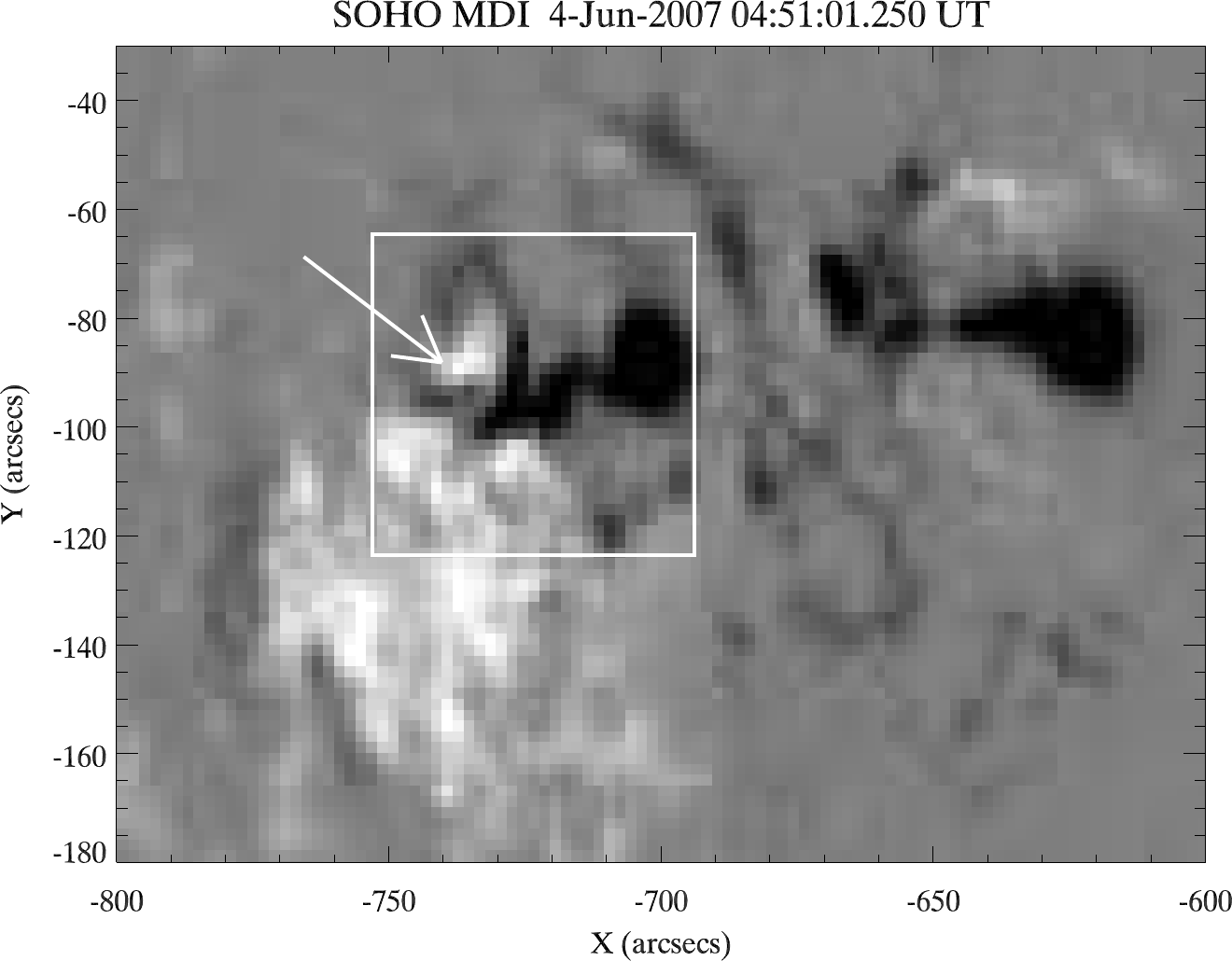}

\includegraphics[width=5.5cm]{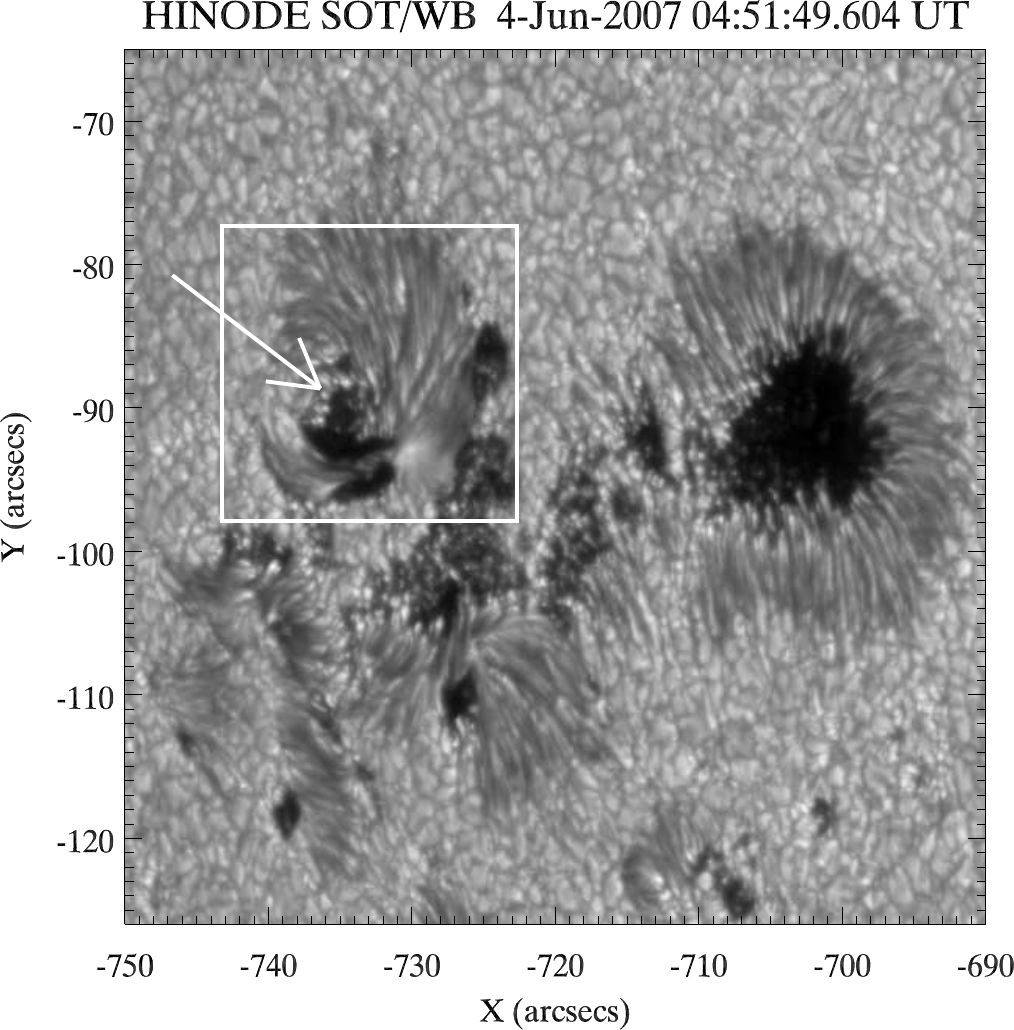}
\includegraphics[width=5.1cm]{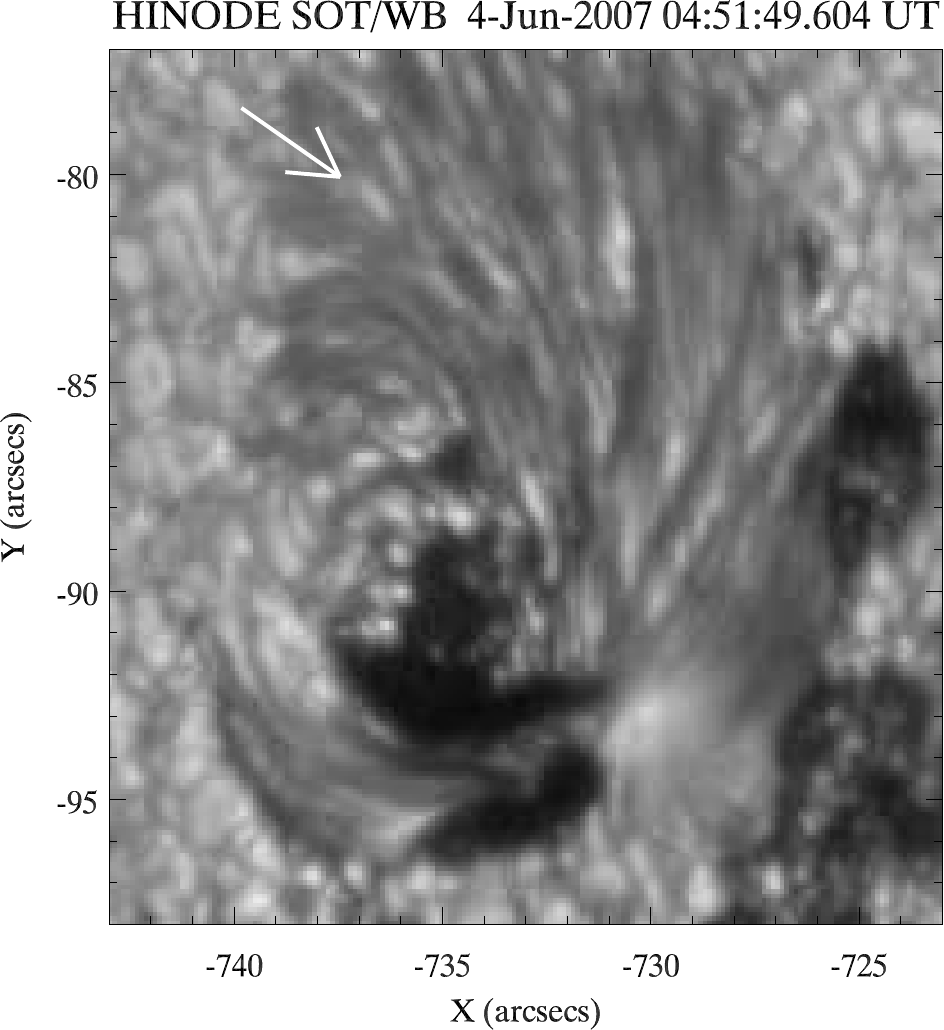}
\caption{Top: SOHO/MDI image of active region NOAA 10960 on 04 June 2007. The positive-polarity sunspot indicated by arrow plays an important role in triggering the M8.9/3B solar flare. The enlarged view of the sunspot group of the active region as indicated by a box in SOHO/MDI image, is shown in the SOT/blue continuum (4504 \AA) image (bottom-left panel). A more closer view of the positive-polarity sunspot as indicated by a box in the SOT/blue continuum image, is also shown in the bottom-right image. Penumbral filaments, twisted in the counterclockwise direction, are clearly evident in this image.}
\label{fig1}
\end{figure}
\inlinecite{nandy2008} has shown the formation, evolution, and ejection of the magnetic flux ropes, which originate in the twisted magnetic structures through the combined action of surface flux transport processes (such as diffusion, meridional circulation and differential rotation). \inlinecite{ishii1998} have investigated a flaring active region NOAA 5395 during March 1989 and found some peculiar vortex-like motion of small, satellite sunspots, which successively emerged from the leading edge of the sunspot group. They propose a schematic model of successive emergence of twisted and winding magnetic-flux loopdf coiling around a trunk of magnetic-flux tube, and concluded the flare triggering due to this emerging flux bundles.  The twisted flux-tube model \cite{amari2000} and flux-injection driven model \cite{krall2001}, both suggest that a twist-enhanced flux rope can play a crucial role in large-scale eruptive events. The primary mechanism for driving such eruptions may be due to the catastrophic loss of  MHD equilibrium \cite{lin2000}. They have suggested that a flux rope is allowed to escape with a fairly small reconnection rate in the vertical current sheet created below. The highly twisted flux tubes store magnetic energy, which is necessary for the heating and particle acceleration during the solar eruptions. \inlinecite{wang2002} have found a rapid disappearance of a sunspot associated with the M2.4 flare from NOAA AR 9830 on 20 February 2002 with hard X-ray sources located near this disappeared sunspot.  \inlinecite{ishii2000} have also pointed out that the occurrence of high flare activity is restricted to the location and the time at which the strongly twisted magnetic flux ropes emerged to the photosphere in the long lived and large active regions, {\it e.g.}, NOAA AR 4201.

The observational evidences of twisted helical structure are less abundant in the solar atmosphere. However, they may be an efficient mechanism for the triggering of solar eruptive phenomena  e.g. flares and coronal mass ejections (CMEs).  \inlinecite{gary2004} and \inlinecite{liu2003}  have observed the helical magnetic flux tubes with multiple turns, which were associated with double flares and CMEs. This observational evidence of the activation of helical flux tubes, and thus associated destabilization of large-scale magnetic fields of active region, may be important clues for the energy build-up processes of solar flares. However, we do not have sufficient understanding and observational signature of the twisted flux ropes and their evolution from the sub-photospheric level into the corona. There are a few observational signatures related to the generation of the twist in the solar filaments, which causes the disruption of their stable magnetic field configuration and generates solar eruptive events ({\it e.g.}, \opencite{liu2009}; \opencite{williams2005} and references cited there). \inlinecite{rust2005} have also found evidence of sigmoids in the solar corona which were governed and energized by the magnetic twist, without any large-scale destabilization of magnetic fields and associated eruptions from the Sun. \inlinecite{gerrard2002} have found that the foot-point twisting motion may also generate the twisting in the active-region loopdf which trigger flare event after their reconnection with the surrounding opposite-polarity field lines.

AR 10960 shows the successive activation of helical twisted magnetic structures on 4 June 2007, which do not cause any eruption. Recently, evidence of the kink instability has been found in the right-handed twisted loop, which causes the B5.0 class flare in this AR during \,04:40--04:51\, UT \cite{sri2010}. In the present paper, we study the M8.9/3B flare event of the same active region NOAA 10960 during \,05:06--05:16\, UT on 4 June 2007 using multi-wavelength observations. We find   rare observational evidence of the activation of helical-twisted magnetic structure in the active region, which may produce the M-class flare.  In Section 2, we present multi-wavelength observations of AR 10960 and the associated flare. In later sections, we present the discussion and conclusions.

\section{Multiwavelength Observations of NOAA 10960 and Associated M8.9/3B Flare}
The flare event was observed by various space based instruments namely
 SOHO/MDI, {\it Hinode}/SOT, TRACE and STEREO/SECCHI. The top panel of Figure 1 displays the SOHO/MDI image of the active region NOAA 10960 on 4 June 2007 before this flare activity. The active region is located nearby the eastern limb at S09E50 showing a $\beta\gamma\delta$ configuration. This active region produced ten M-class flares during its passage across the solar disk. However, this active region was very poor in CME production, and only two M-class flares were associated with CMEs \cite{yashiro2008}. In the present study, the M8.9/3B flare was triggered without any CME eruption observed on 4 June 2007. The positive-polarity sunspot is indicated by an arrow, which plays an important role in triggering the M8.9/3B solar flare. The enlarged view of the sunspot group, as indicated by a box on SOHO/MDI image, is shown in the SOT/blue continuum image (4504 \AA) (bottom-left panel). The closer view of the positive-polarity sunspot, as indicated by a box in SOT/blue continuum image, is also shown in the bottom-right panel. The penumbral filaments, twisted in the counterclockwise direction, are clearly evident in this image. 

 According to the GOES soft X-ray flux profiles in the \,0.5--4\, \AA \ and \,1--8\, \AA \ wavelength bands, the M8.9 flare starts at 05:06 UT, reaches maximum at 05:13 UT, and ends at 05:16 UT (Figure 9). This flare shows an impulsive rise for a short duration during the above-mentioned time period and then a gradual decay for a long time until nearly 06:45 UT. However, a small B5.0 class flare was also observed well before this flare event during \,04:40--04:51\, UT, which seems to be a precursor for M8.9 flare.  According to the Solar Geophysical Data (SGD), the M-class flare is classified as 3B class in H$\alpha$, where, the flare starts at 05:05 UT, peaks at  05:14 UT and ends at 06:42 UT.

\subsection{{\it Hinode}/SOT Observations} 
\begin{figure} 
\centerline{
\hspace*{-0.015\textwidth}
\includegraphics[width=0.25\textwidth]{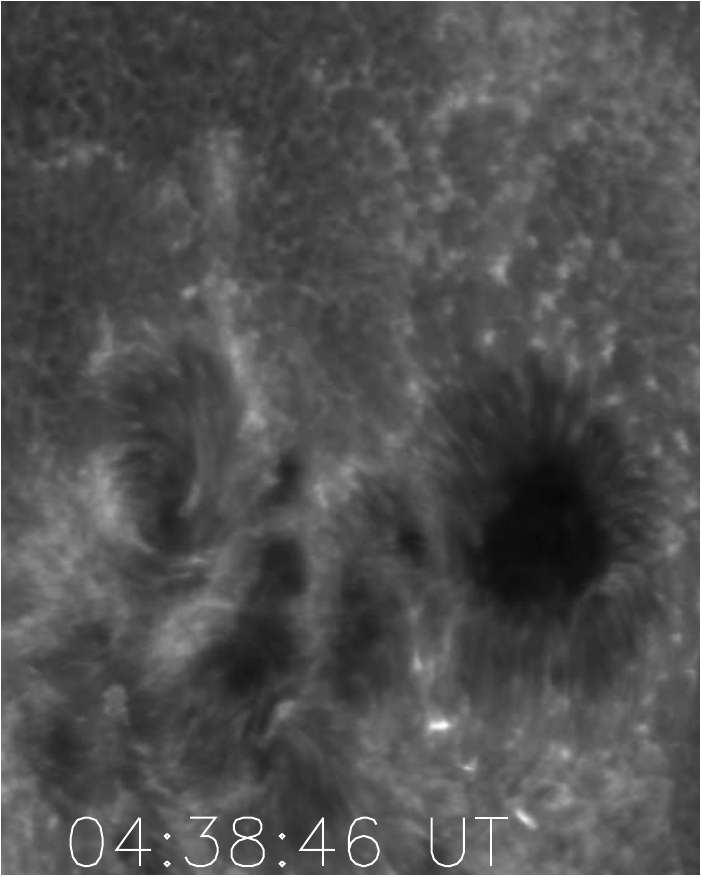}
\hspace*{-0.015\textwidth}
\includegraphics[width=0.25\textwidth]{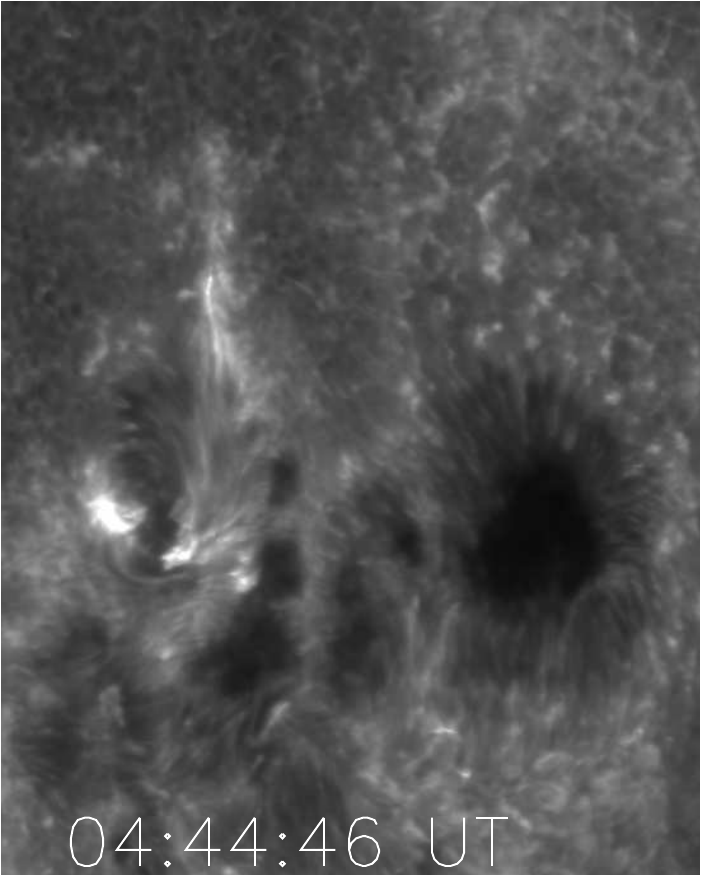}
\hspace*{-0.015\textwidth}
\includegraphics[width=0.25\textwidth]{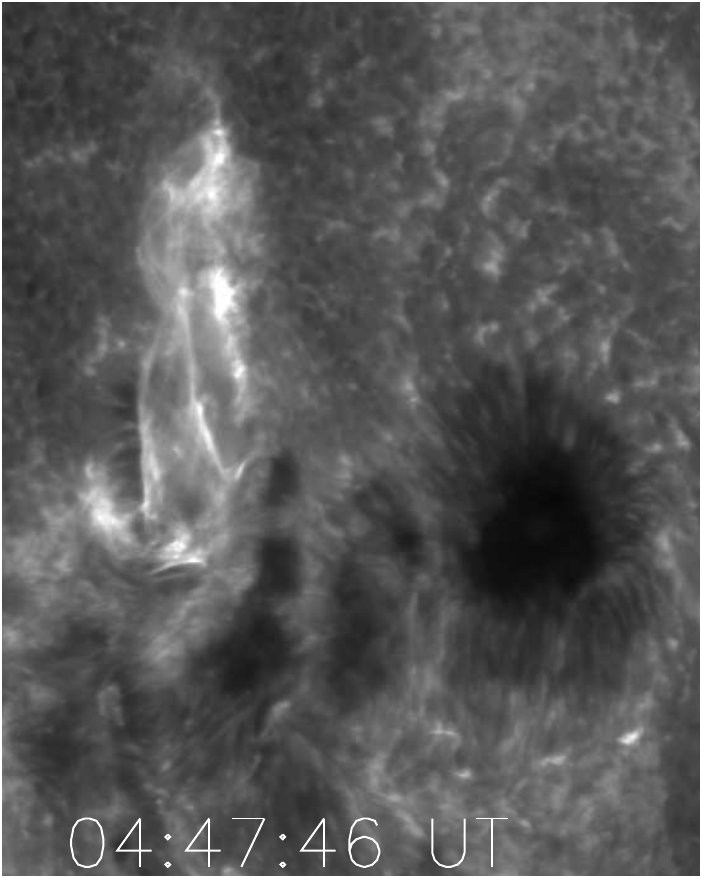}
\hspace*{-0.015\textwidth}
\includegraphics[width=0.25\textwidth]{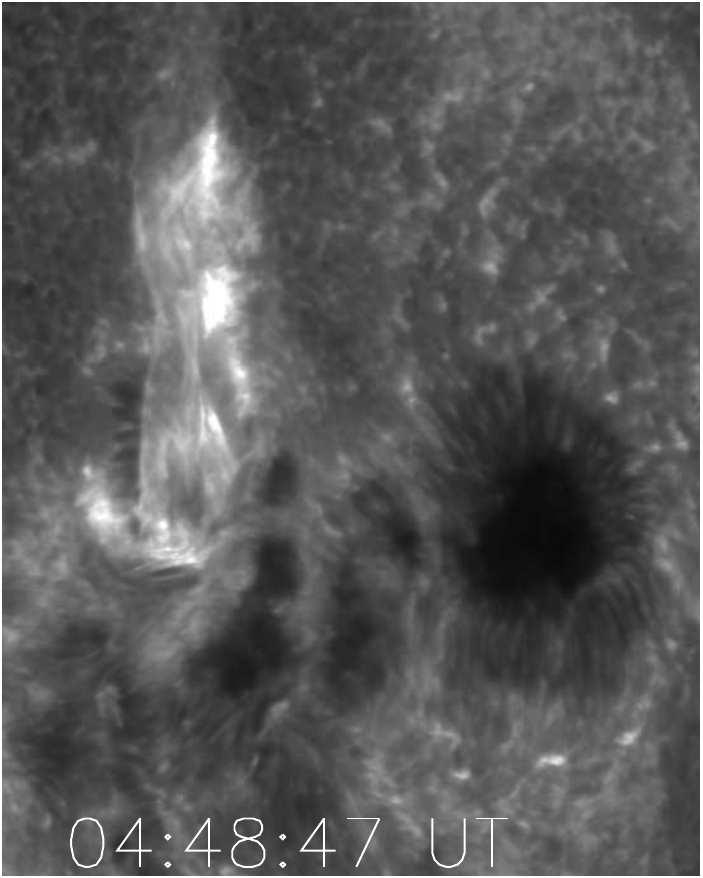}
\hspace*{-0.015\textwidth}
}
\centerline{
\hspace*{-0.015\textwidth}
\includegraphics[width=0.25\textwidth]{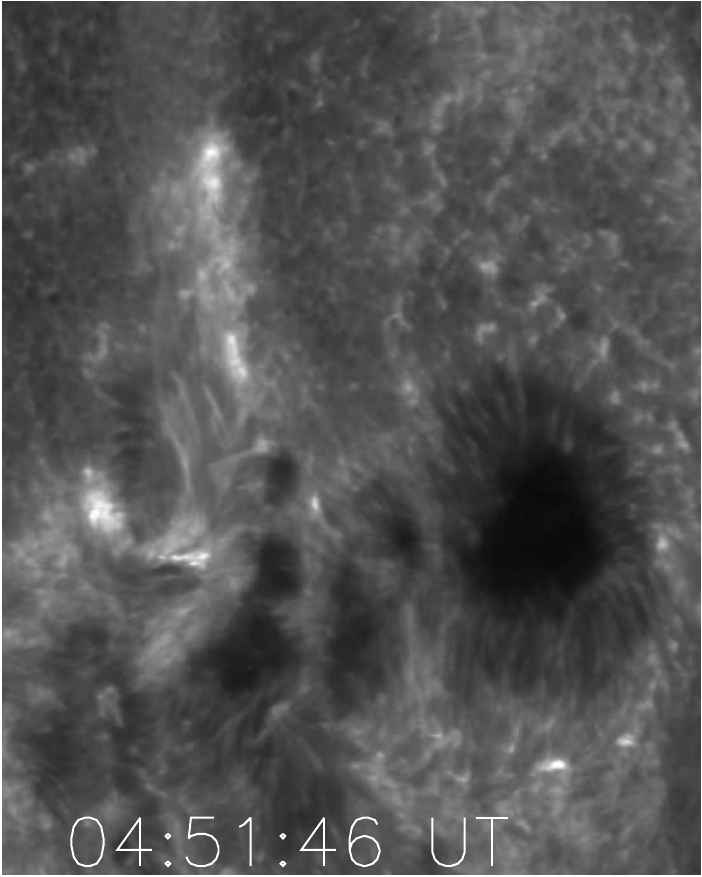}
\hspace*{-0.015\textwidth}
\includegraphics[width=0.25\textwidth]{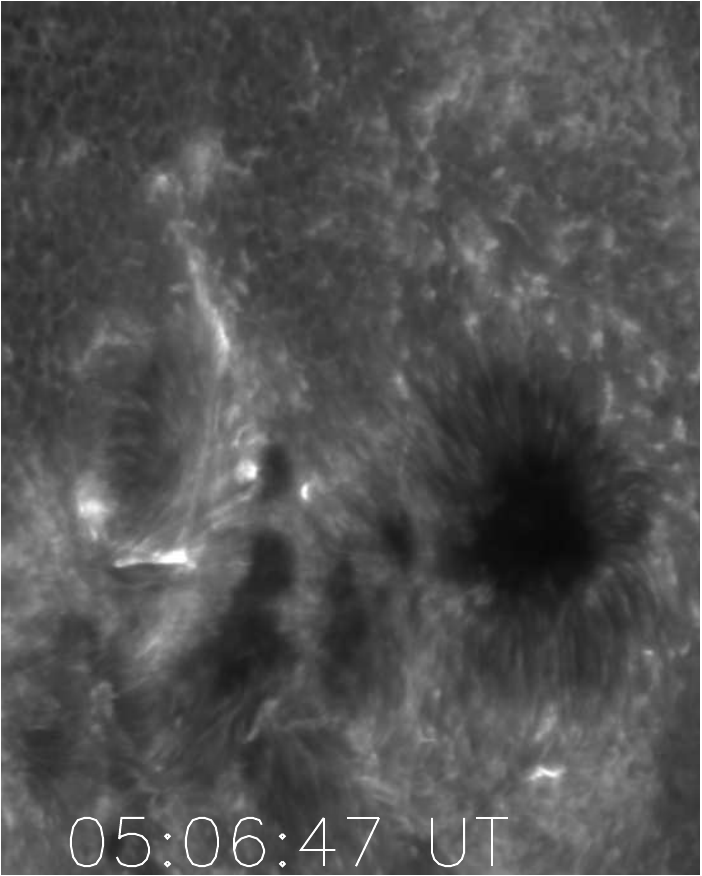}
\hspace*{-0.015\textwidth}
\includegraphics[width=0.25\textwidth]{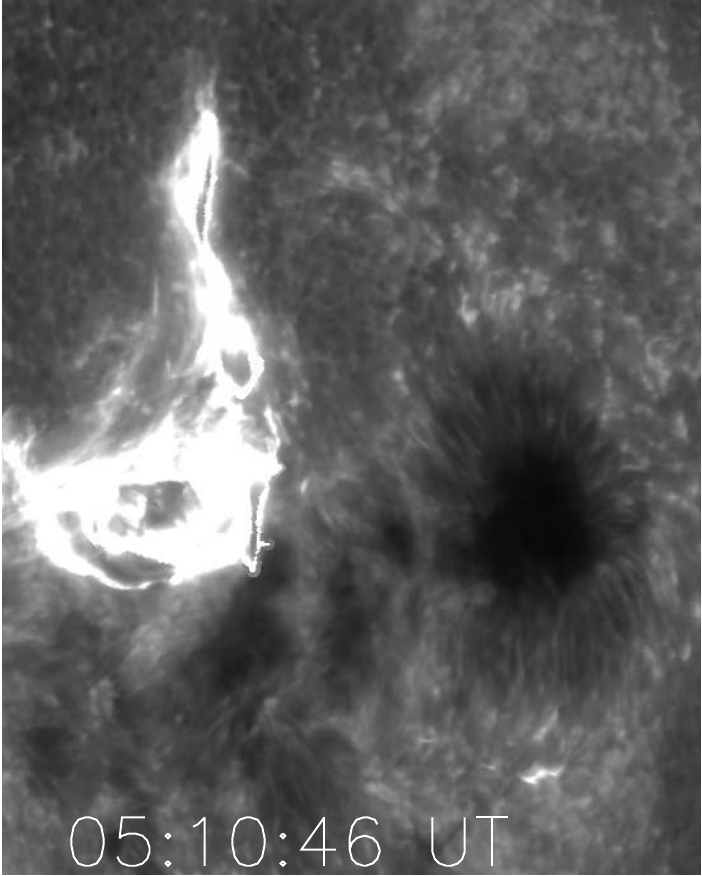}
\hspace*{-0.015\textwidth}
\includegraphics[width=0.25\textwidth]{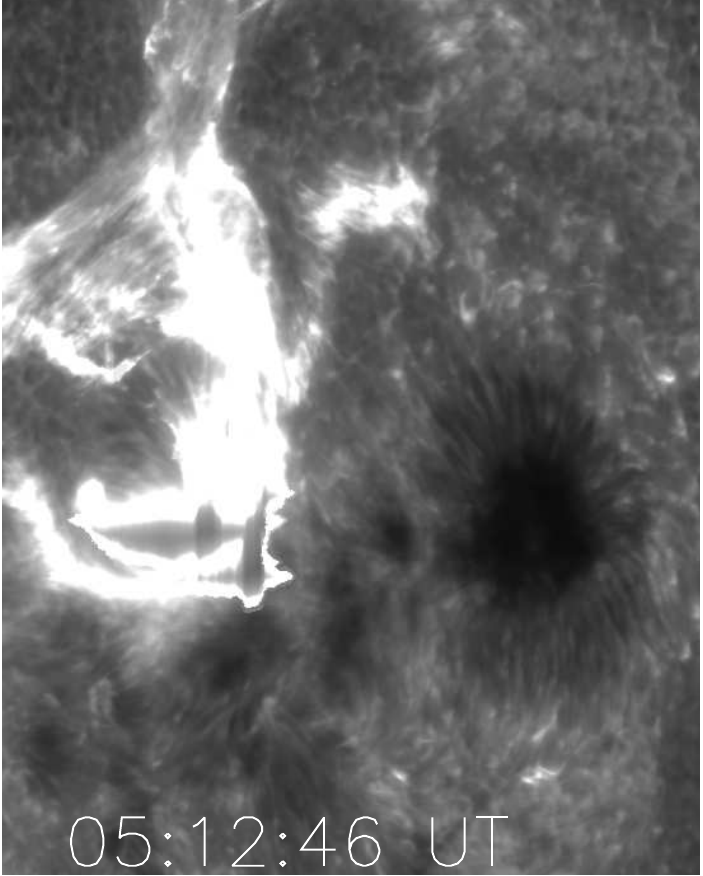}
\hspace*{-0.015\textwidth}
}
\centerline{
\hspace*{-0.015\textwidth}
\includegraphics[width=0.25\textwidth]{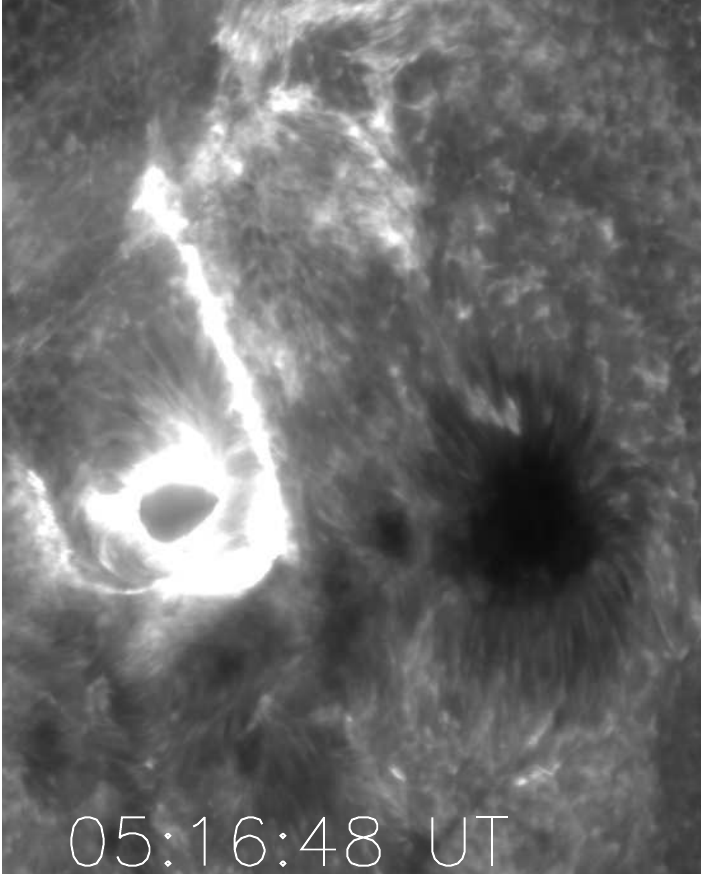}
\hspace*{-0.015\textwidth}
\includegraphics[width=0.25\textwidth]{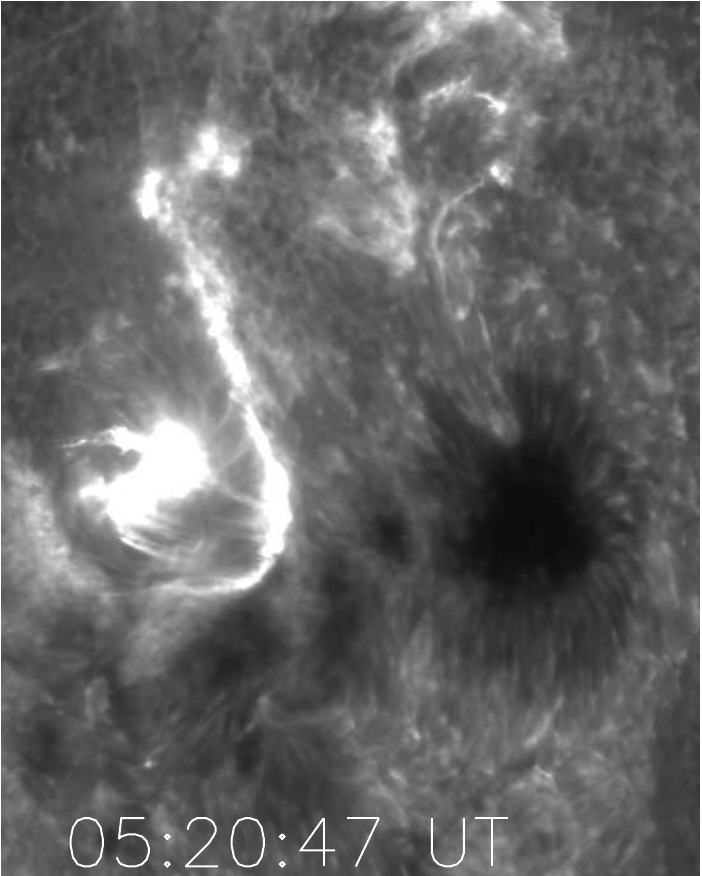}
\hspace*{-0.015\textwidth}
\includegraphics[width=0.25\textwidth]{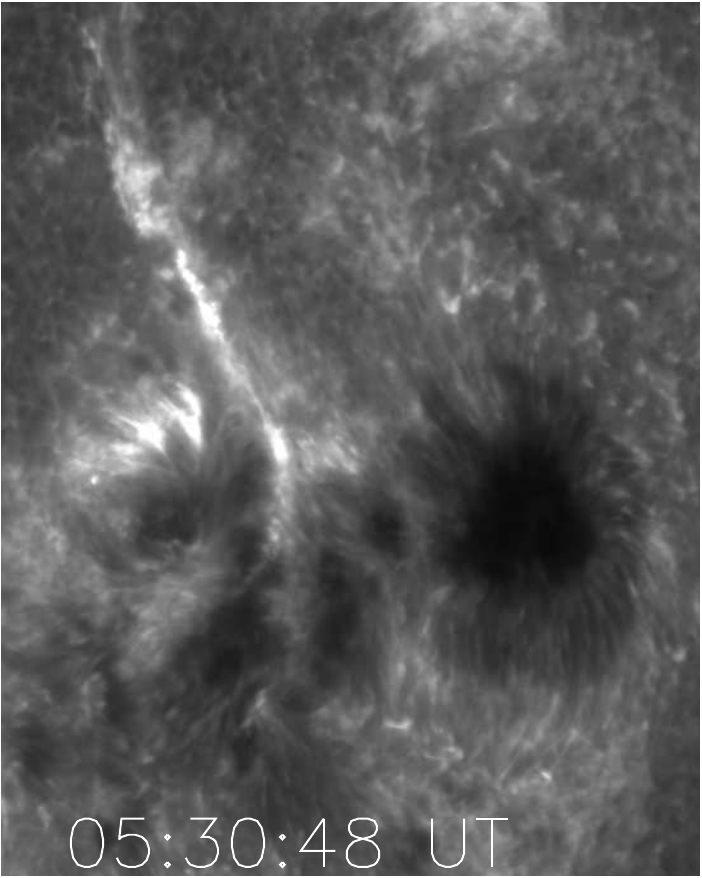}
\hspace*{-0.015\textwidth}
\includegraphics[width=0.25\textwidth]{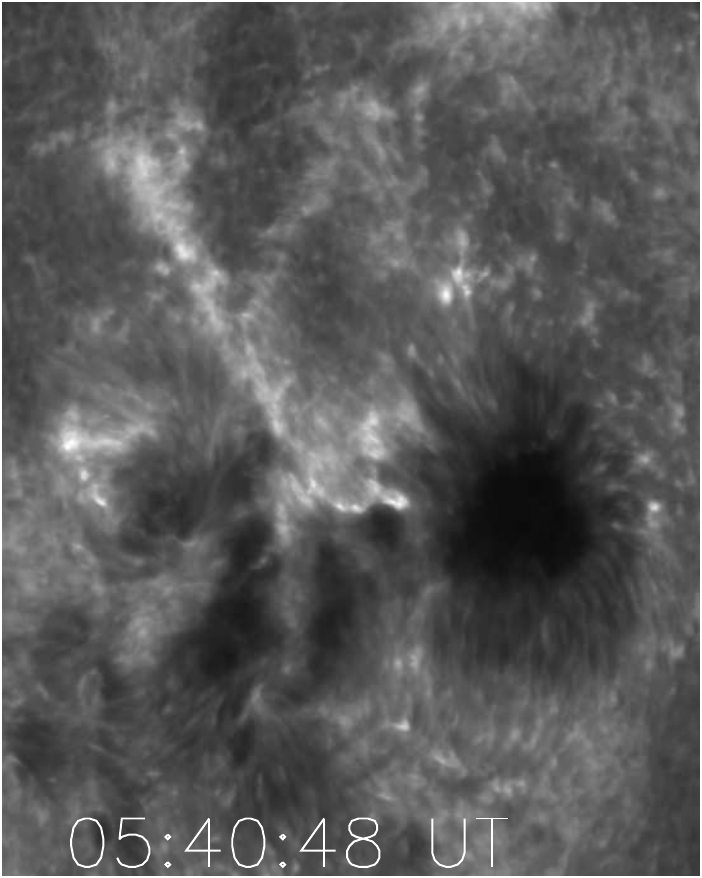}
\hspace*{-0.015\textwidth}
}
\caption{{\it Hinode}/SOT Ca {\sc ii} H 3968 \AA \ images showing the successive activation of helical twists and corresponding brightening above the positive-polarity sunspot on 4 June 2007. Secondary helical twist has been activated at $\approx$05:08 UT and causes the maximum of the M8.9/3B class flare brightening in the chromosphere at $\approx$05:13 UT. The size of each image is 60$^{\prime\prime}$$\times$75$^{\prime\prime}$.}
\label{fig2}
\end{figure}
The high-resolution filtergrams of a flaring region in NOAA 10960 were obtained by the 50 cm {\it Solar Optical Telescope} (SOT) onboard the {\it Hinode} spacecraft. We use the SOT/Ca {\sc ii} H 3968 \AA \  chromospheric images  at a cadence of $\approx$ one minute, with a spatial resolution of 0.1$^{\prime\prime}$ per pixel \cite{tsuneta2008}. We also use SOT/blue-continuum (4504 \AA) temporal image data to examine the changes and evolution of the positive-polarity sunspot which is the key place at the centre of the active region where the flare activity occured. The {\it Hinode} data are calibrated and analysed using standard IDL routines in the SolarSoft {\sc (ssw)} package. Figure 2 displays the selected chromospheric temporal images of the active region NOAA 10960 in Ca {\sc ii} H line. These images show the significant changes in the chromosphere before the flare. The image at 04:42 UT shows the two bright points at the two opposite edges of an umbral-bridge structure. After that, a twisted, bright structure appears near the umbral-bridge from the southern part of the sunspot. It is at $\approx$29,000 km (in projection) from the sunspot location. This structure was visible for nearly \,six--seven\, minutes during \,04:45--04:51\, UT and then after it fades out against the chromosphere \cite{sri2010}. 
  

After this first episode, two other bright points were observed at the same site before the initiation of the M8.9 flare (refer to the image at 05:06 UT). After this, the flare starts and it covers the  full sunspot at the maximum phase.  The ``S''-shaped single ribbon, formed at 05:16 UT, was observed until 05:26 UT. The flare continues until 05:30 UT. Therefore, at the same location, near the positive-polarity sunspot, the twisted structure showed two successive activations in association with the energy build-up and release processes in the AR 10960. It seems that the activation of secondary twist at $\approx$05:08 UT plays a significant role in the triggering of M8.9/3B solar flare.

\subsection{TRACE and STEREO/SECCHI/EUVI Observations}

\begin{figure} 
\centerline{
\hspace*{-0.02\textwidth}
\includegraphics[width=0.4\textwidth]{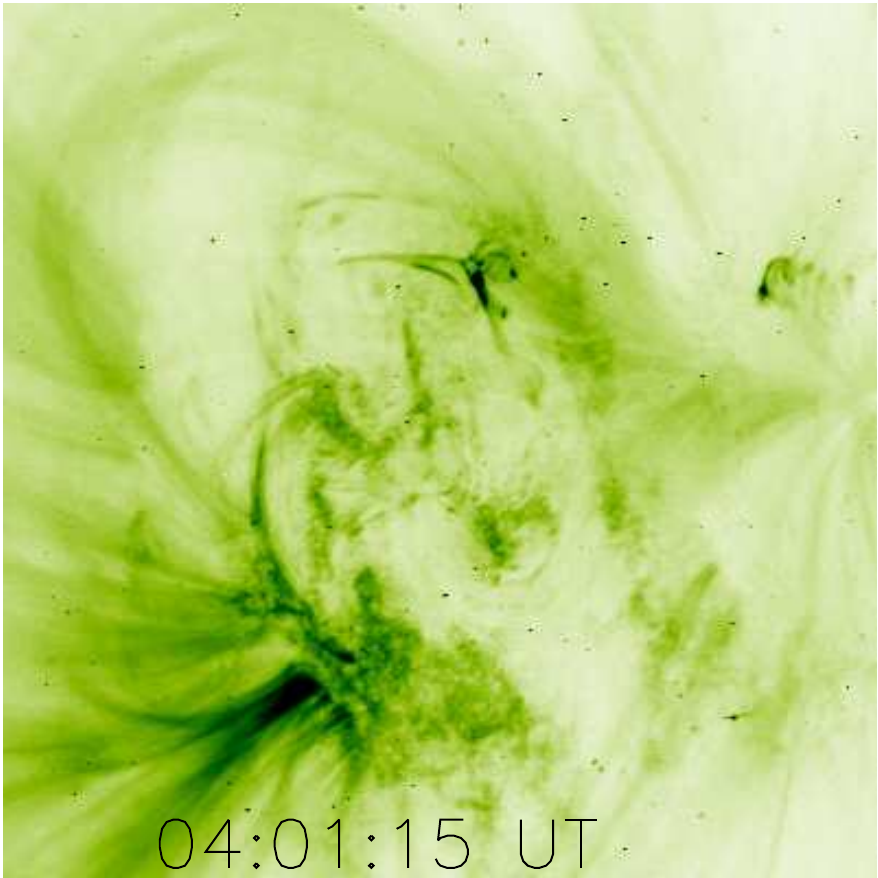}
\hspace*{-0.02\textwidth}
\includegraphics[width=0.4\textwidth]{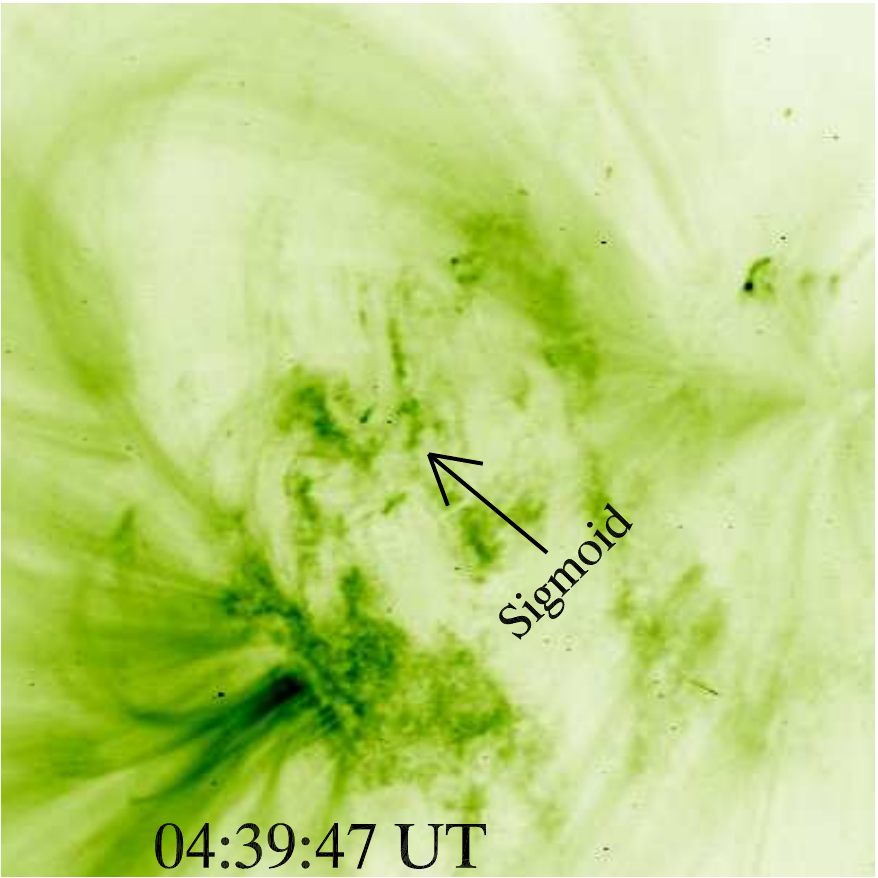}
\hspace*{-0.02\textwidth}
}
\centerline{
\hspace*{-0.02\textwidth}
\includegraphics[width=0.4\textwidth]{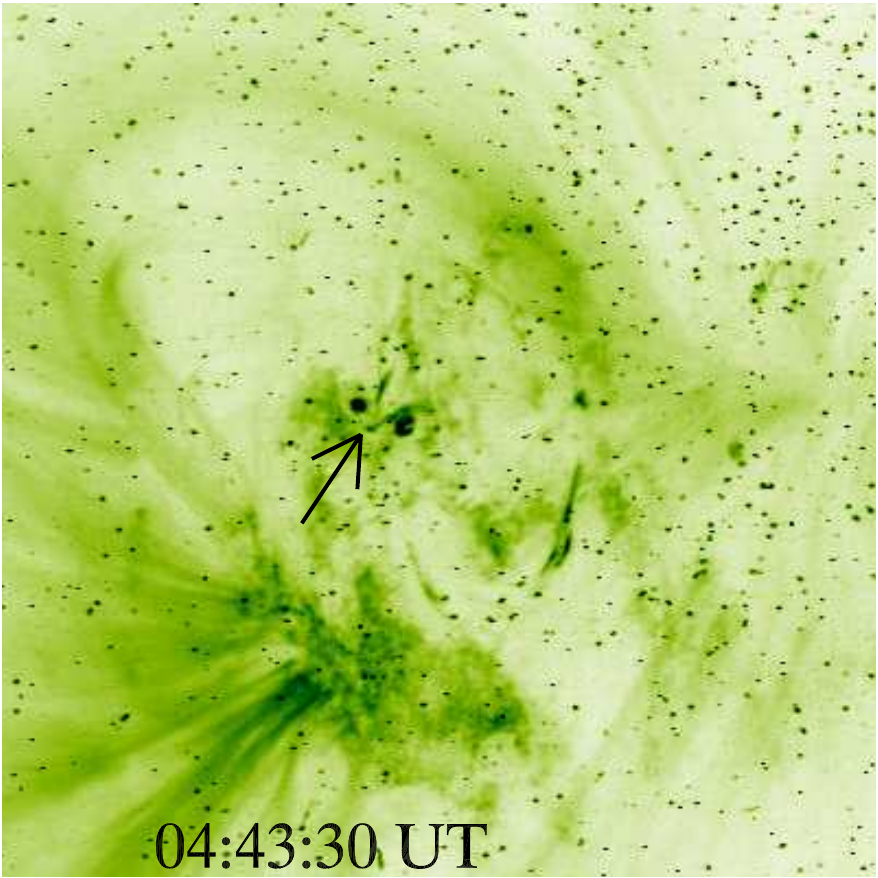}
\hspace*{-0.02\textwidth}
\includegraphics[width=0.4\textwidth]{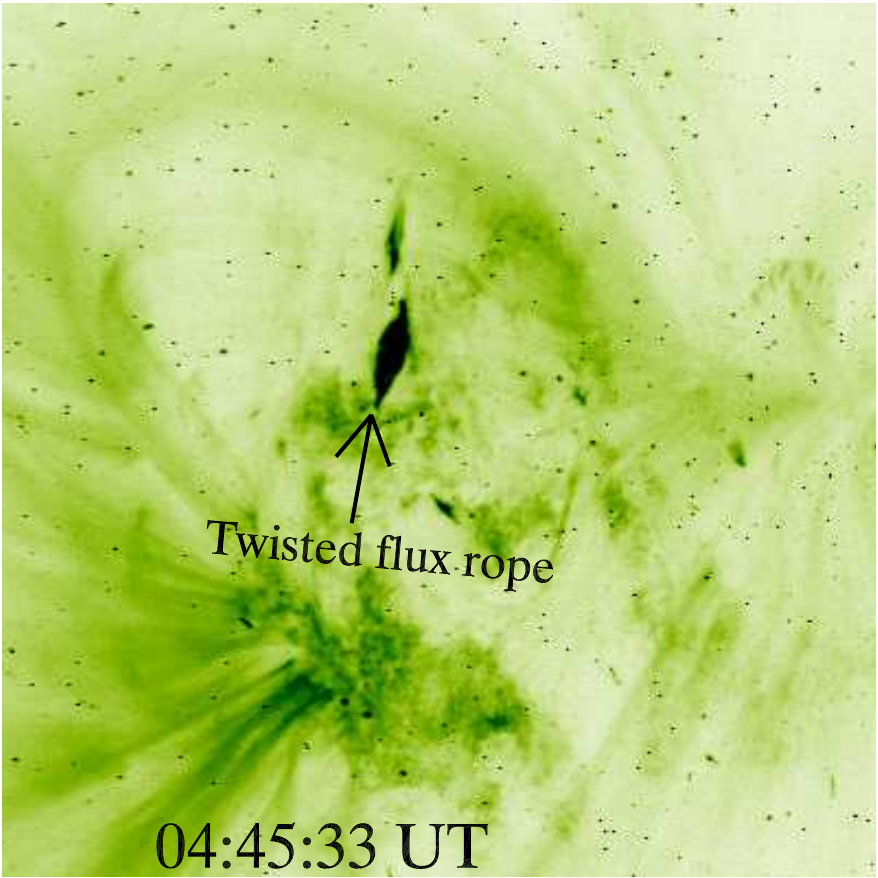}
\hspace*{-0.02\textwidth}
}

\centerline{
\hspace*{-0.02\textwidth}
\includegraphics[width=0.4\textwidth]{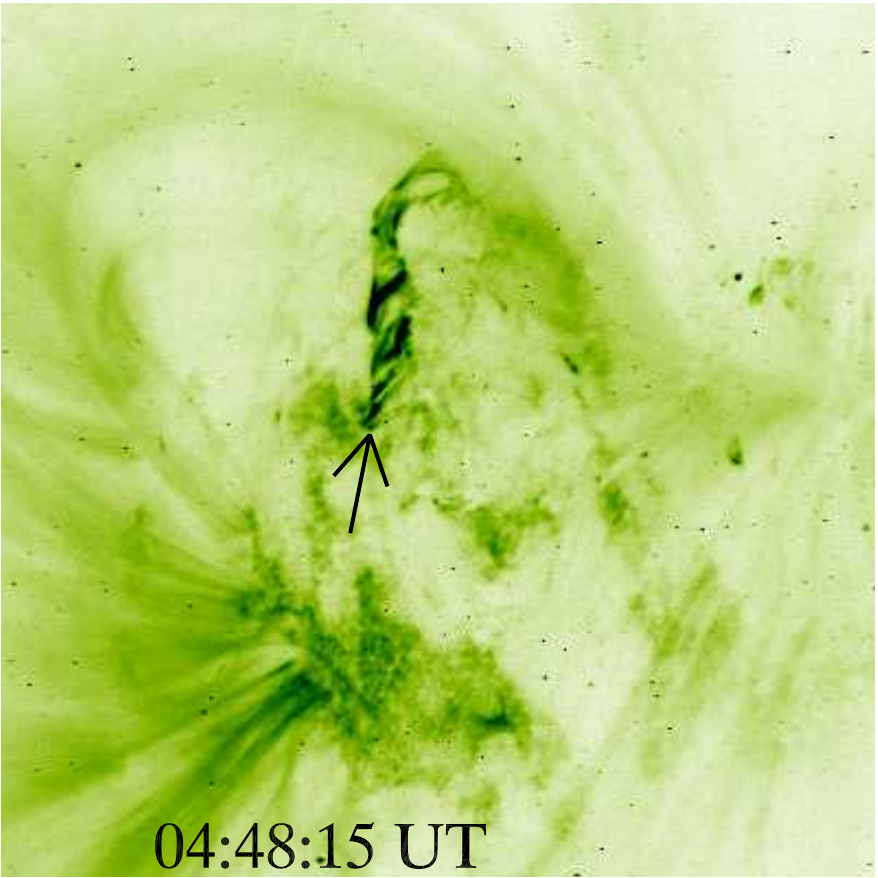}
\hspace*{-0.02\textwidth}
\includegraphics[width=0.4\textwidth]{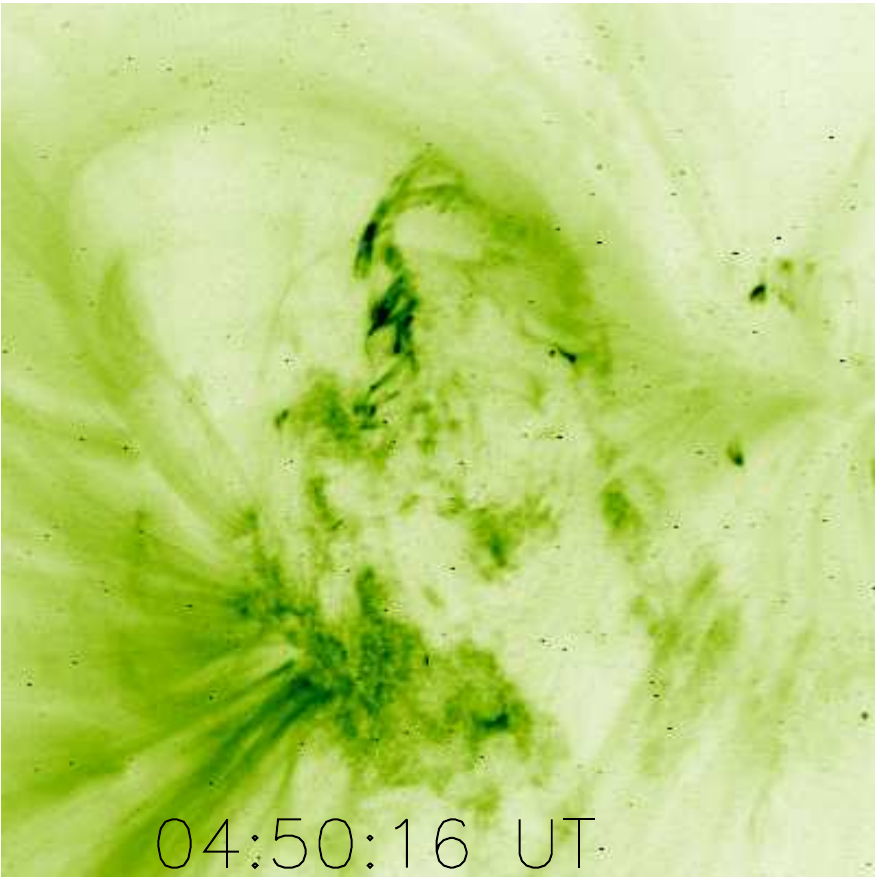}
\hspace*{-0.02\textwidth}
}
\centerline{
\hspace*{-0.02\textwidth}
\includegraphics[width=0.4\textwidth]{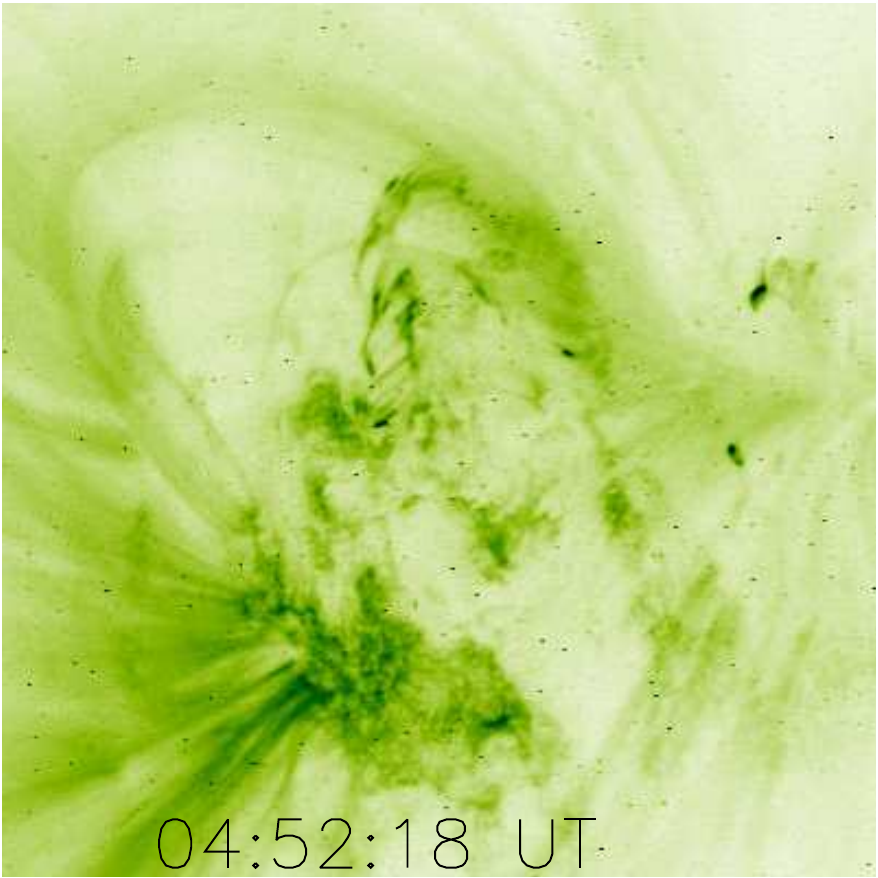}
\hspace*{-0.02\textwidth}
\includegraphics[width=0.4\textwidth]{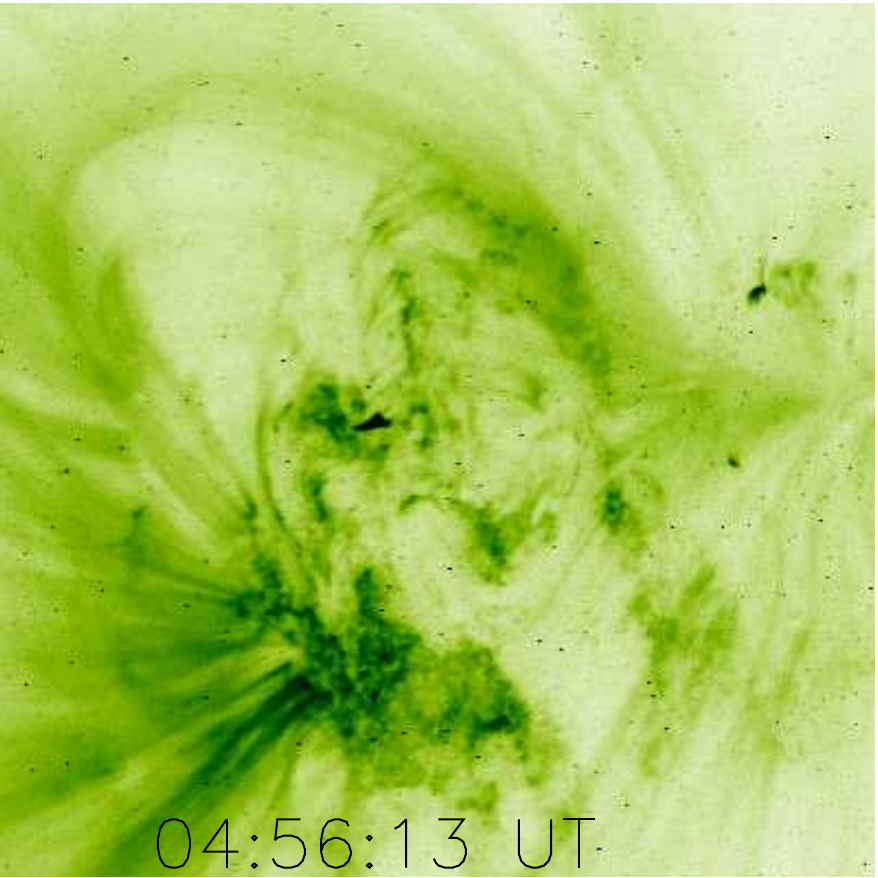}
\hspace*{-0.02\textwidth}
}
\caption{TRACE 171 \AA \ EUV images (in reversed colors) showing the temporal changes in the magnetic field configuration before the initiation of M-class flare. The size of each image is 200$^{\prime\prime}$$\times$200$^{\prime\prime}$.}
\label{fig3}
\end{figure}
We use TRACE 171 \AA \ (Fe {\sc ix}) EUV images to study the dynamics of the flaring active region and its response in the corona before the M-class flare event. This wavelength corresponds to 1.3 MK plasma. The image size is 1024$\times$1024 pixels with the resolution of 0.5$^{\prime\prime}$ per pixel, and the cadence is $\approx$ one minute. We have used the standard IDL routines available in the SolarSoft library for cleaning and co-aligning the images. Figure 4 displays the selected TRACE coronal images before the flare during \,04:00--04:56\, UT. The careful  investigation of the TRACE movie shows the plasma flow from the flare site along the two close and smaller loopdf located within the big loop (see the image at 04:01:15 UT). The image at 04:39 UT shows the S-shaped sigmoid structure (indicated by an arrow) at the active-region centre.  At 04:43 UT, we observe two bright points near the sigmoid structure, which is also evident in SOT/Ca {\sc ii} temporal image data. Therefore,  the first helical twist has appeared in the loop system associated with this particular positive-polarity sunspot and also causes the brightening in the plasma there (see also the first panel of Figure 3). This sunspot is connected with the negative-polarity sunspot by very faint loop system in which the twist and brighting are activated, and their configuration is also highly changeable with time \cite{sri2010}. The projected lower-bound speed of the activation of this twist is $\approx$200 km s$^{-1}$. It spreads to the maximum distance of $\approx$43,500 km from the sunspot (see the image at 04:48 UT). It was visible for nearly \,seven--eight\, minutes. After that, the structure fades out in to the coronal background. TRACE observations of the activation of this primary helical twist correlate nicely with the SOT observations.  However, the length of the structure is a little less in the SOT images in comparison to the TRACE observations. This is due to the smaller field of view of SOT (see the image at 04:48 UT in both sets), which only captured the partial field of view of the helical twisted loop system in the chromosphere. This structure, which is fully observed by TRACE, does not fit in the field of view of the SOT. However, TRACE missed the impulsive phase of this M-class flare. For the impulsive phase of the flare, we use the STEREO-A SECCHI/EUVI observations \cite{wuelser2004}. We use Fe {\sc ix} 171 \AA \ coronal images for the present study. The size of each image is 2048$\times$2048 pixels with 1.6$^{\prime\prime}$ per pixel sampling. We use the standard SECCHI$\_$PREP subroutines for cleaning the images and other standard subroutines available in STEREO  package SolarSoft library. We have used SOHO/MDI images for co-aligning the SECCHI images.

\begin{figure*} 
\centerline{
\hspace*{-0.015\textwidth}
\includegraphics[width=0.33\textwidth]{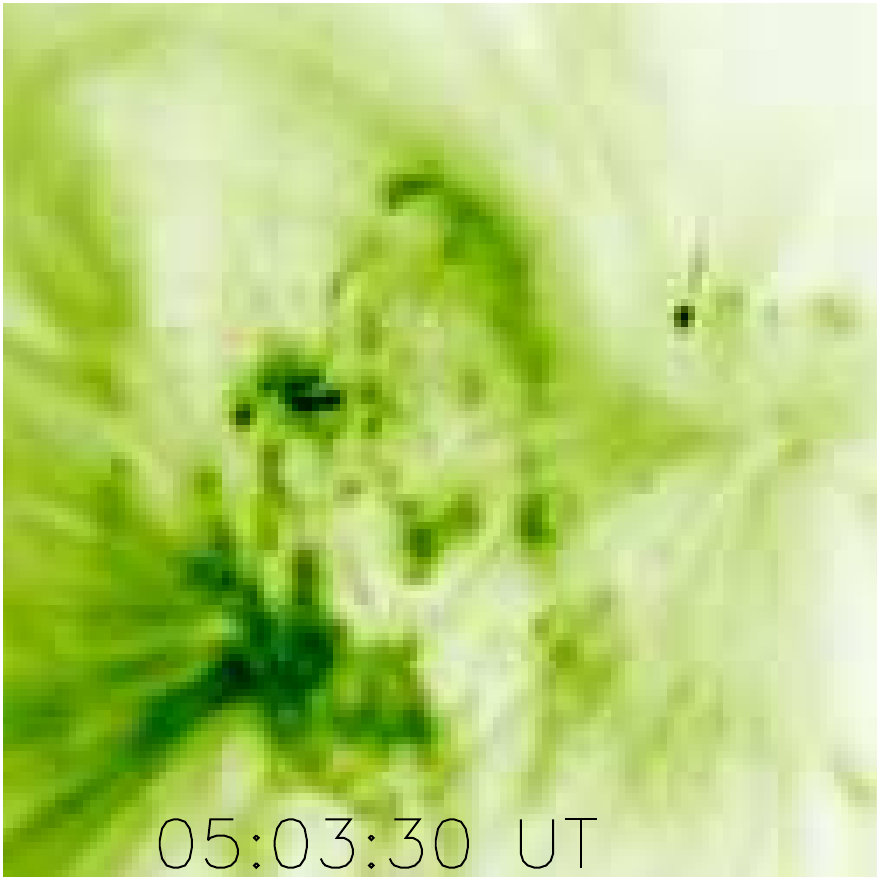}
\hspace*{-0.015\textwidth}
\includegraphics[width=0.33\textwidth]{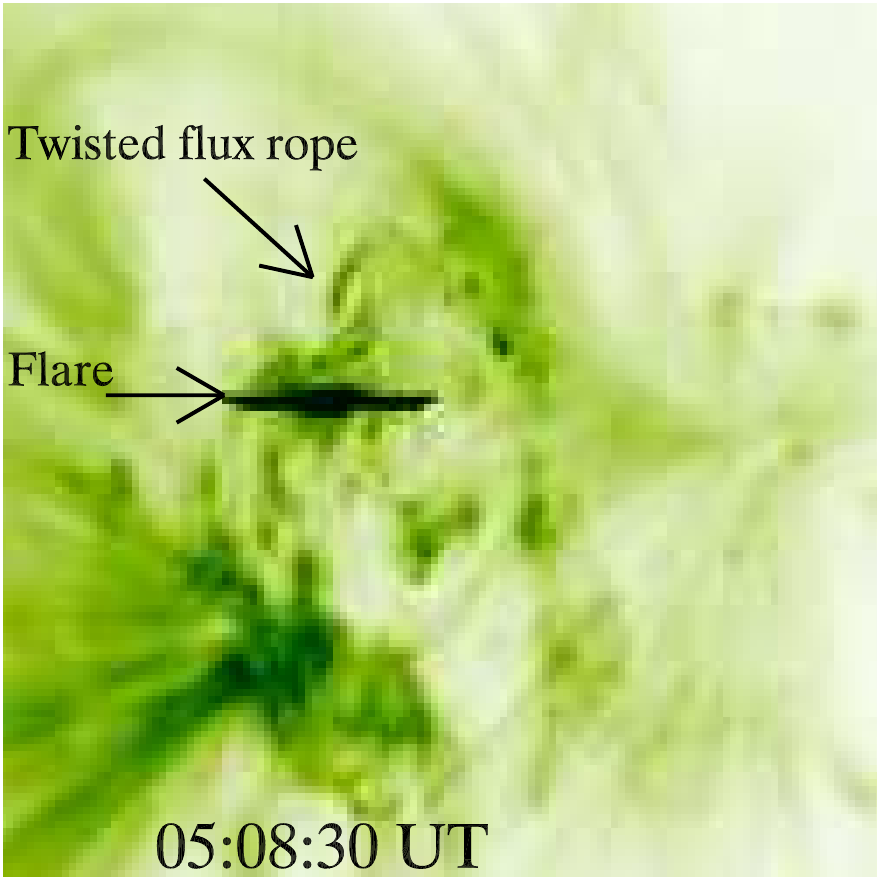}
\hspace*{-0.015\textwidth}
\includegraphics[width=0.33\textwidth]{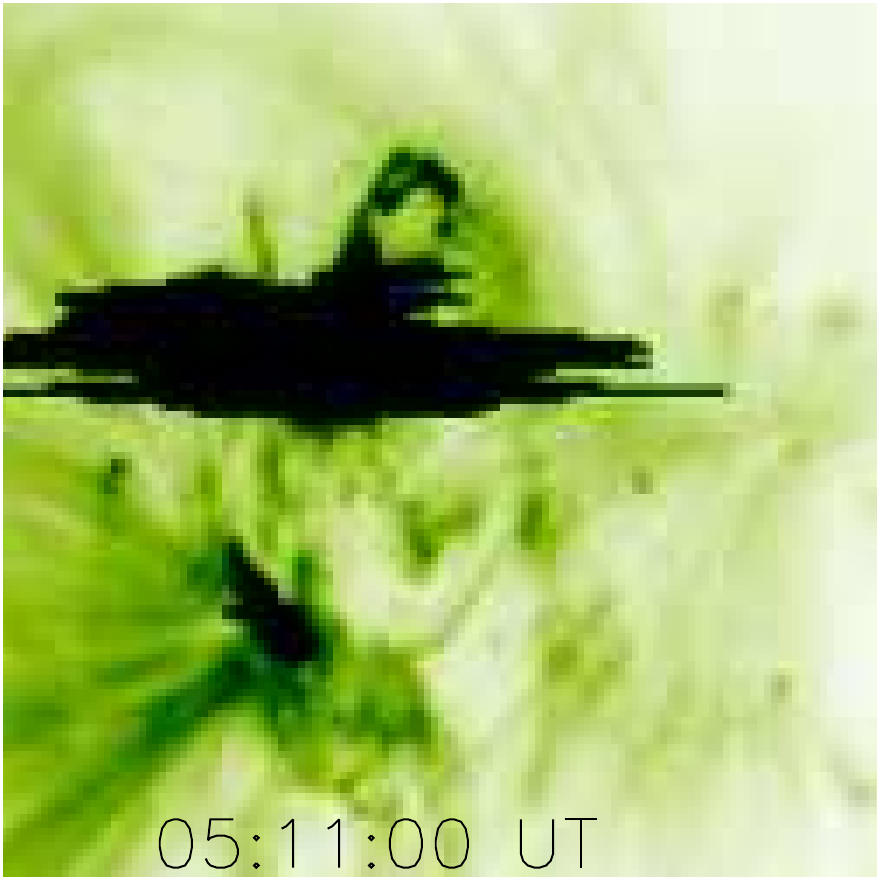}
}
\centerline{
\hspace*{-0.015\textwidth}
\includegraphics[width=0.33\textwidth]{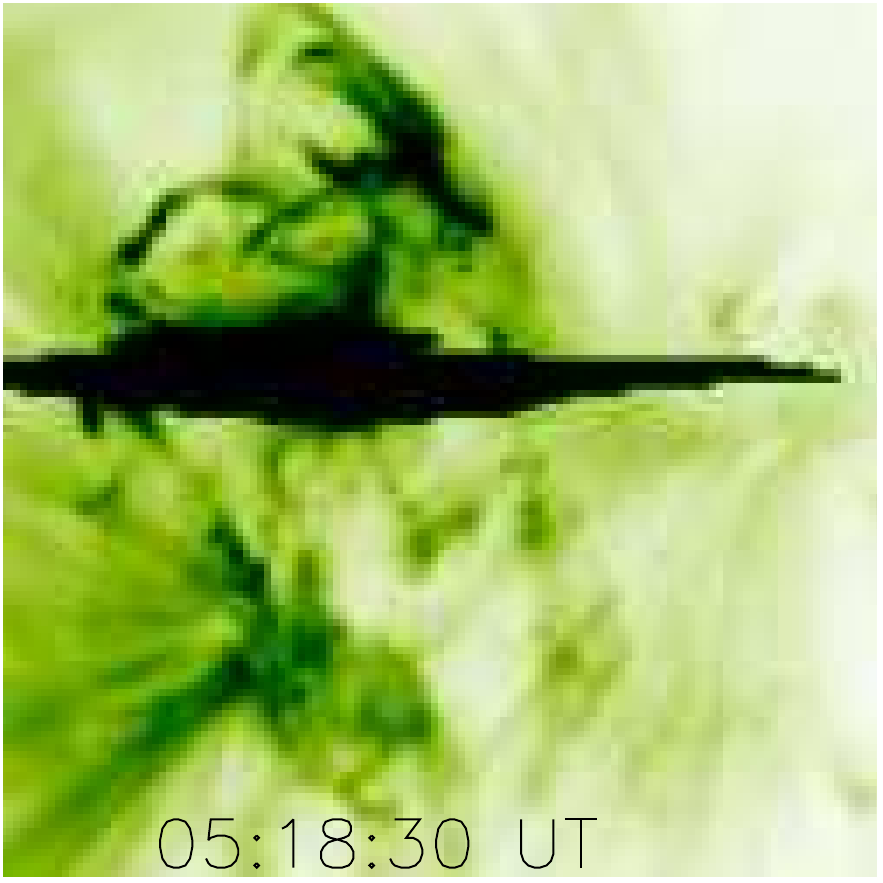}
\hspace*{-0.015\textwidth}
\includegraphics[width=0.33\textwidth]{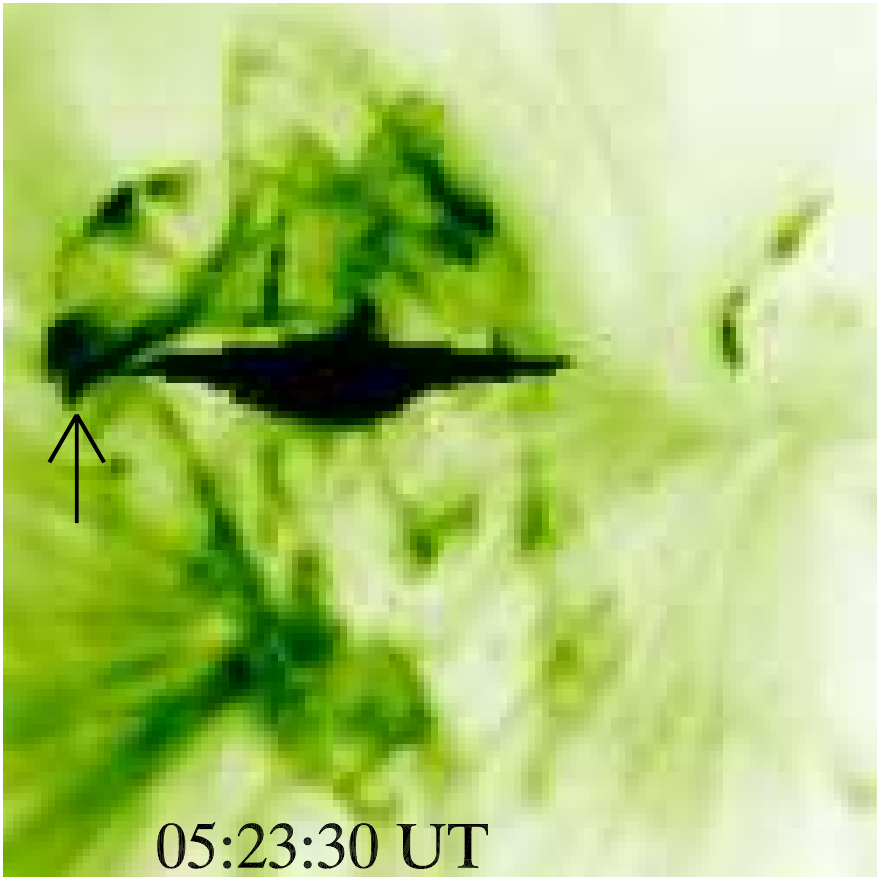}
\hspace*{-0.015\textwidth}
\includegraphics[width=0.33\textwidth]{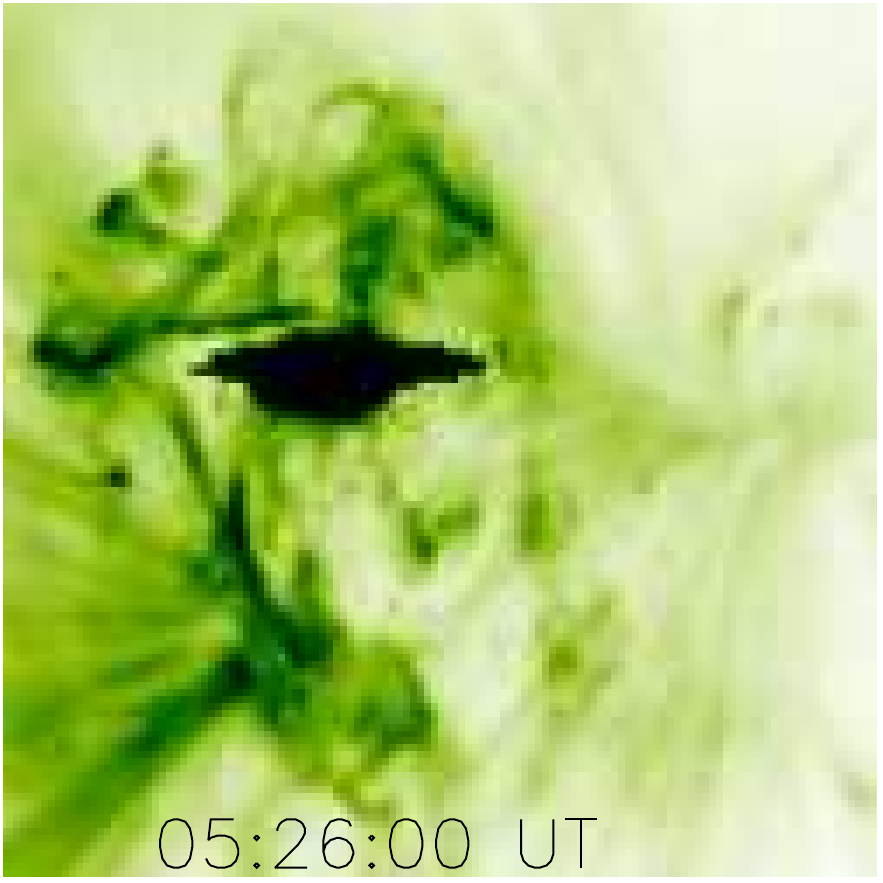}
}
\centerline{
\hspace*{-0.015\textwidth}
\includegraphics[width=0.33\textwidth]{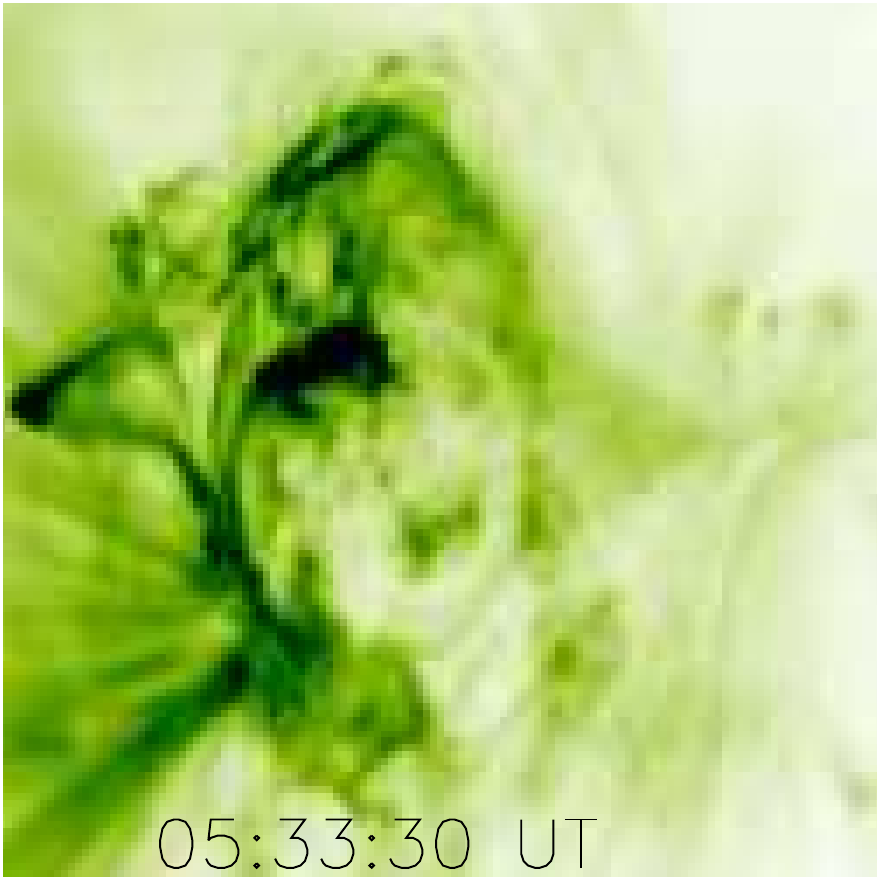}
\hspace*{-0.015\textwidth}
\includegraphics[width=0.33\textwidth]{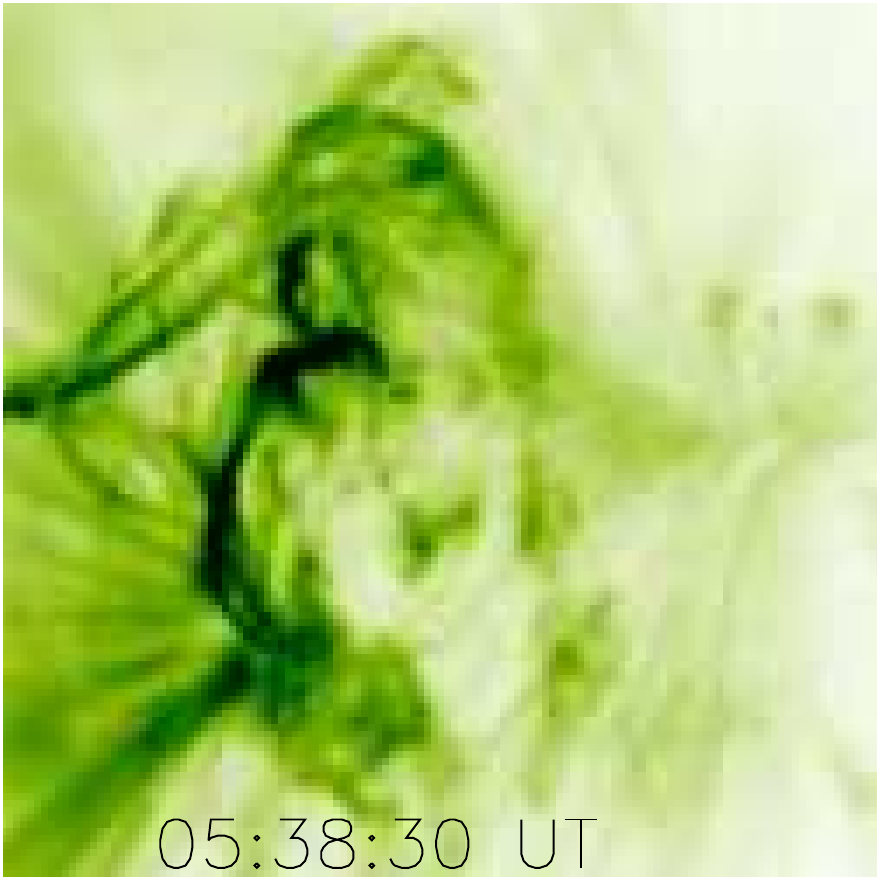}
\hspace*{-0.015\textwidth}
\includegraphics[width=0.33\textwidth]{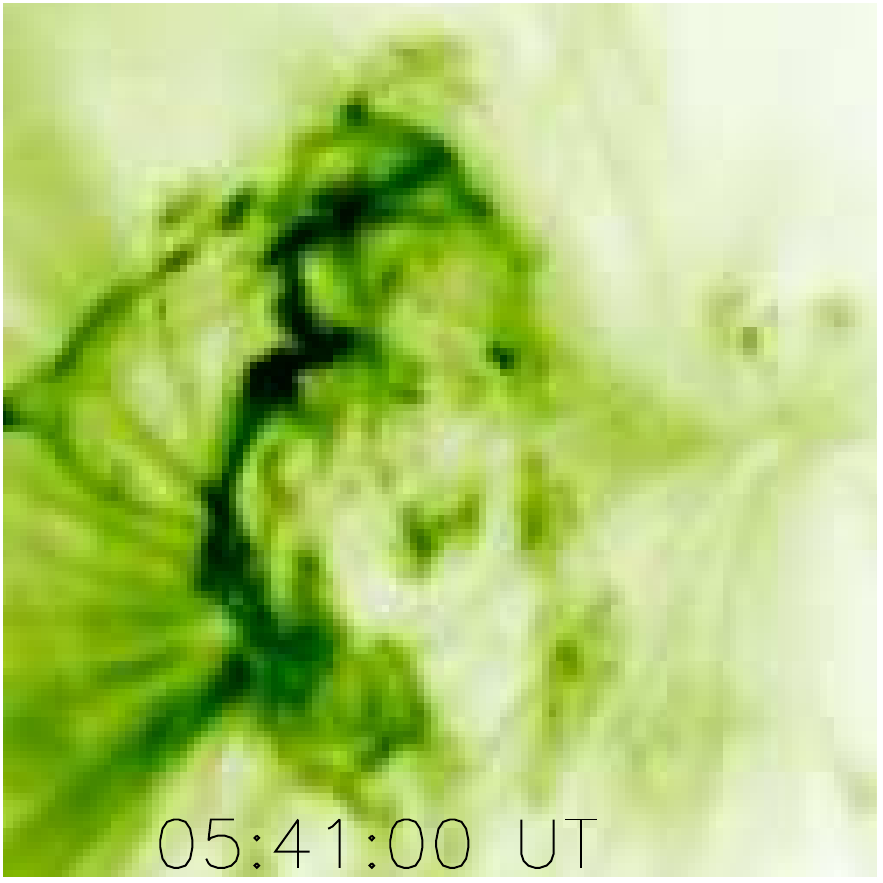}
}
\centerline{
\hspace*{-0.015\textwidth}
\includegraphics[width=0.33\textwidth]{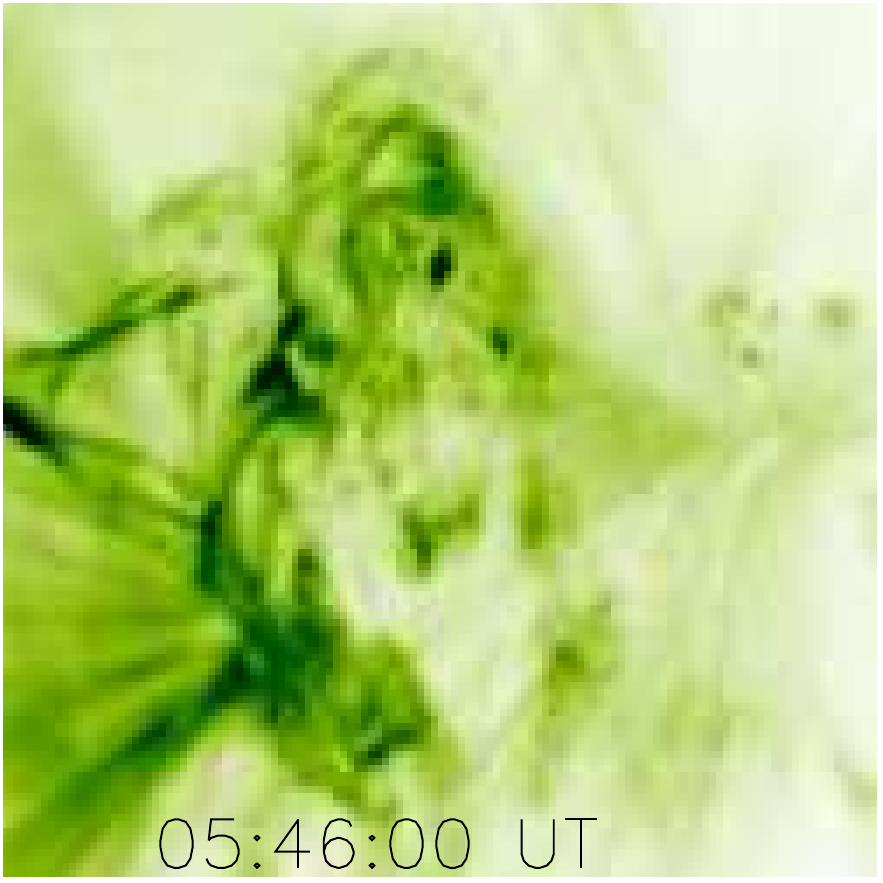}
\hspace*{-0.015\textwidth}
\includegraphics[width=0.33\textwidth]{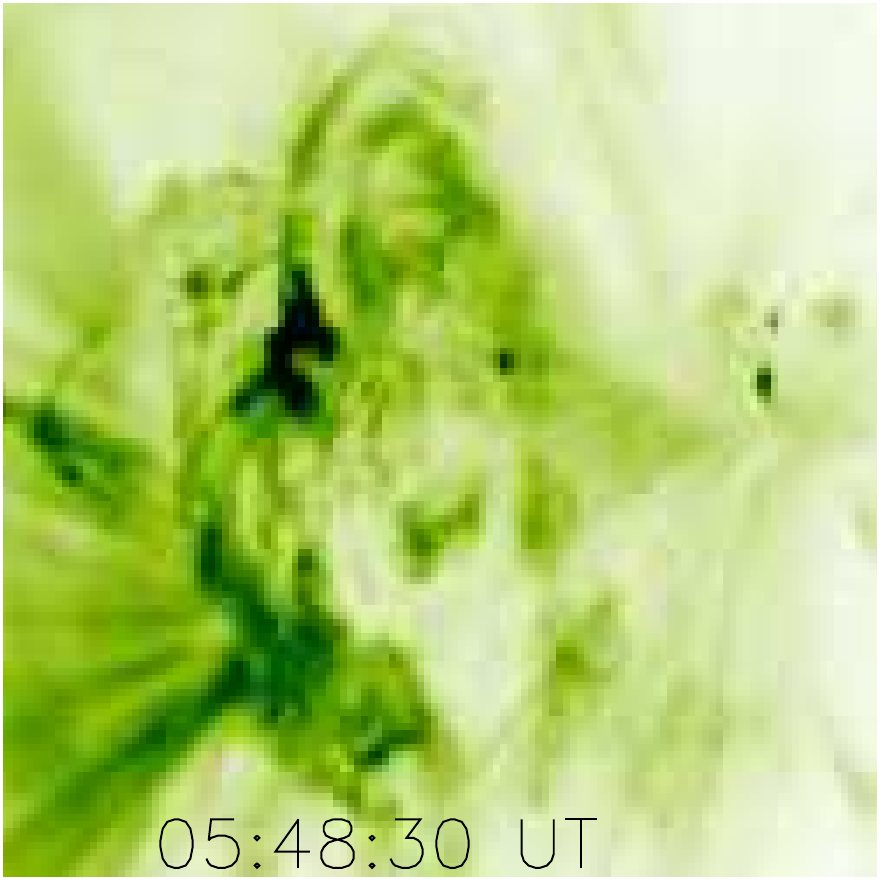}
\hspace*{-0.015\textwidth}
\includegraphics[width=0.33\textwidth]{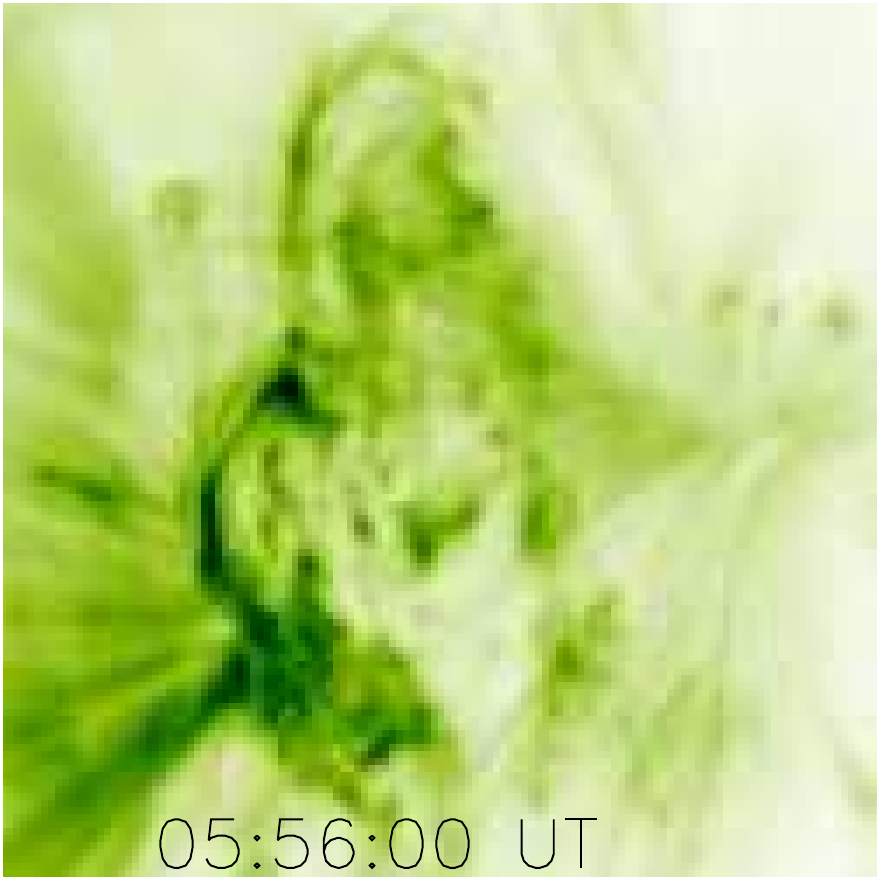}
\hspace*{-0.015\textwidth}
}
\caption{STEREO SECCHI 171 \AA \ images (in reversed colors) showing the temporal changes in the magnetic-field configuration and related M8.9/3B flare event. The size of each image is 200$^{\prime\prime}$$\times$200$^{\prime\prime}$. }
\label{fig4}
\end{figure*}
\begin{figure} 
\centerline{
\hspace*{-0.02\textwidth}
\includegraphics[width=0.4\textwidth]{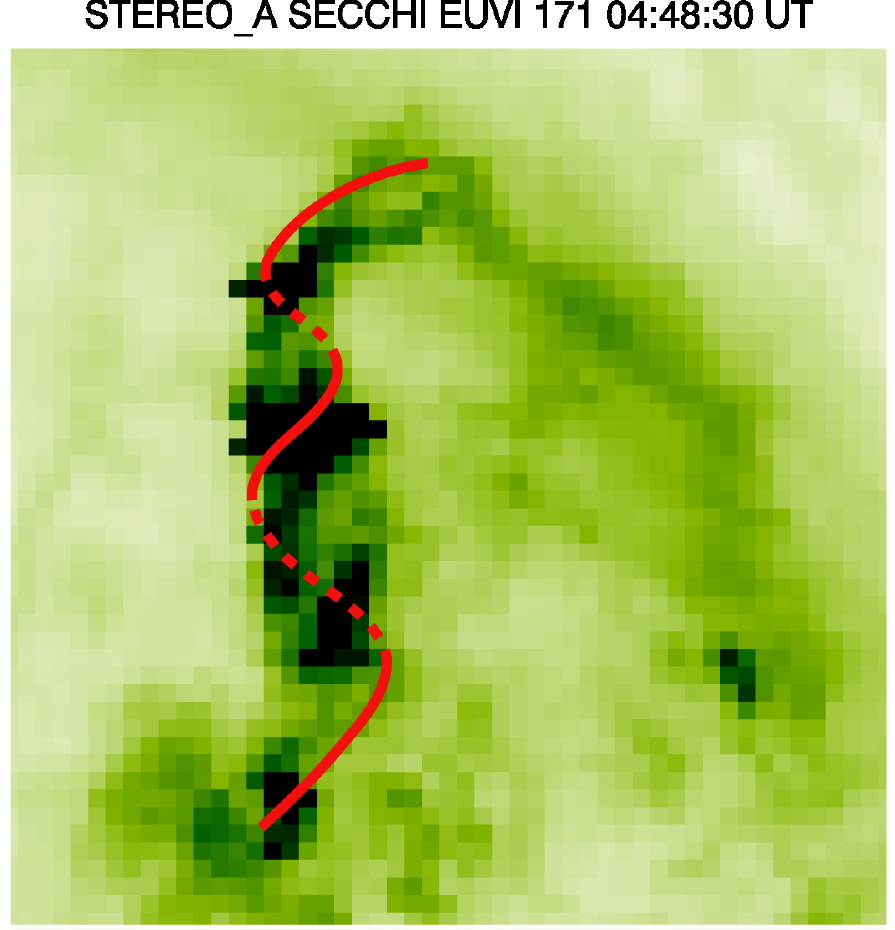}
\hspace*{-0.02\textwidth}
\includegraphics[width=0.4\textwidth]{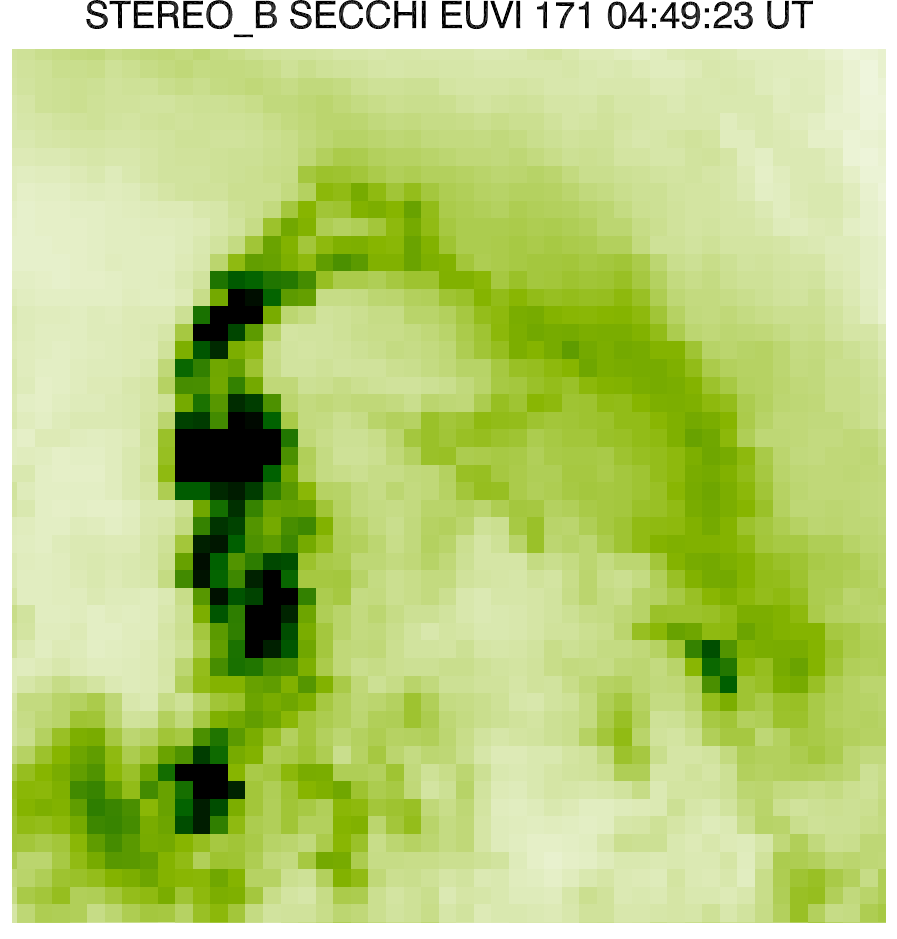}
\hspace*{-0.02\textwidth}}
\centerline{
\includegraphics[width=0.4\textwidth]{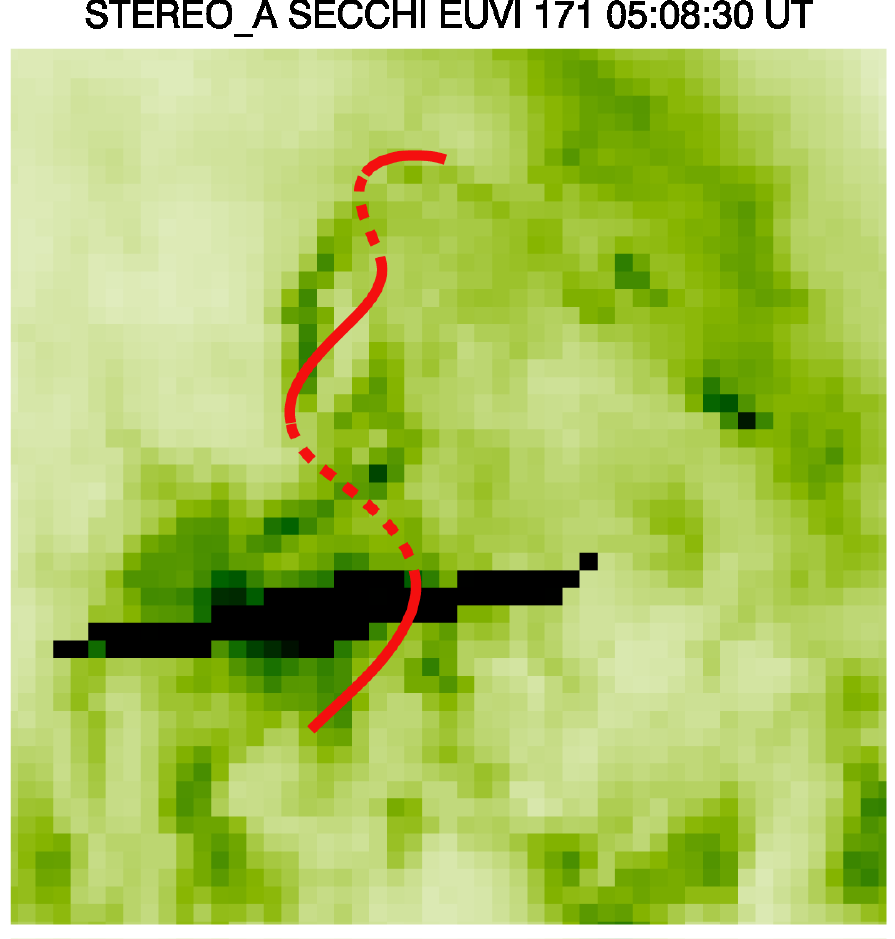}
\hspace*{-0.02\textwidth}
\includegraphics[width=0.4\textwidth]{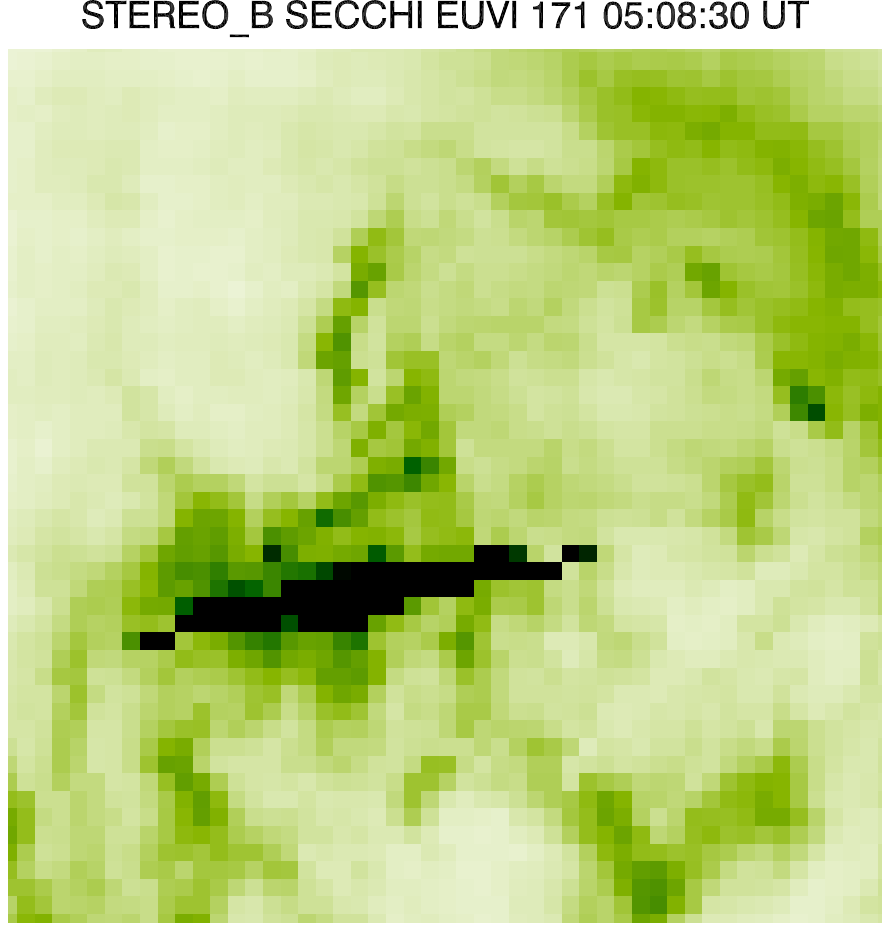}
\hspace*{-0.02\textwidth}
}
\caption{STEREO/SECCHI A and B images of the twisted helical structure. The secondary helical twist with approximately two turns (indicated by red line) has been activated on 05:08 UT just before the maximum of M8.9/3B class flare (bottom panel). The top-left panel shows, for the comparison, three turns (indicated by red line) during the activation of the first helical twist as estimated by Srivastava {\it et al.} (2010). The size of each image is 80$^{\prime\prime}$$\times$80$^{\prime\prime}$.}
\label{fig5}
\end{figure}

\begin{figure} 
\centerline{
\hspace*{-0.02\textwidth}
\includegraphics[width=0.333\textwidth]{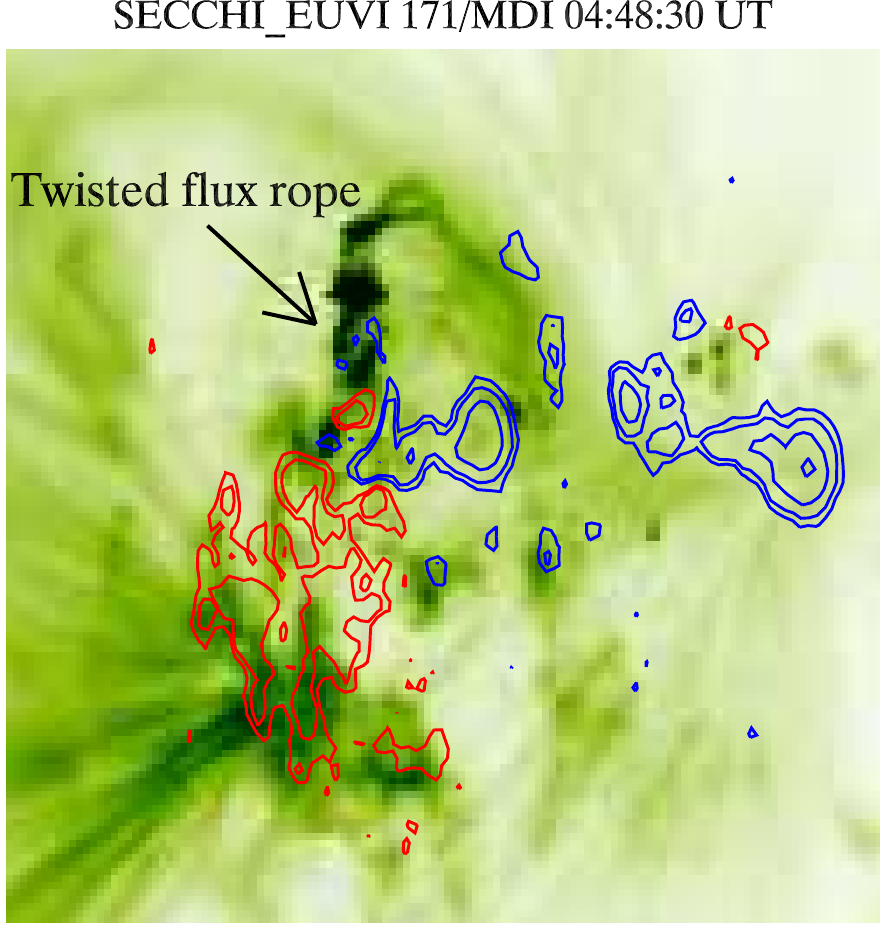}
\hspace*{-0.02\textwidth}
\includegraphics[width=0.333\textwidth]{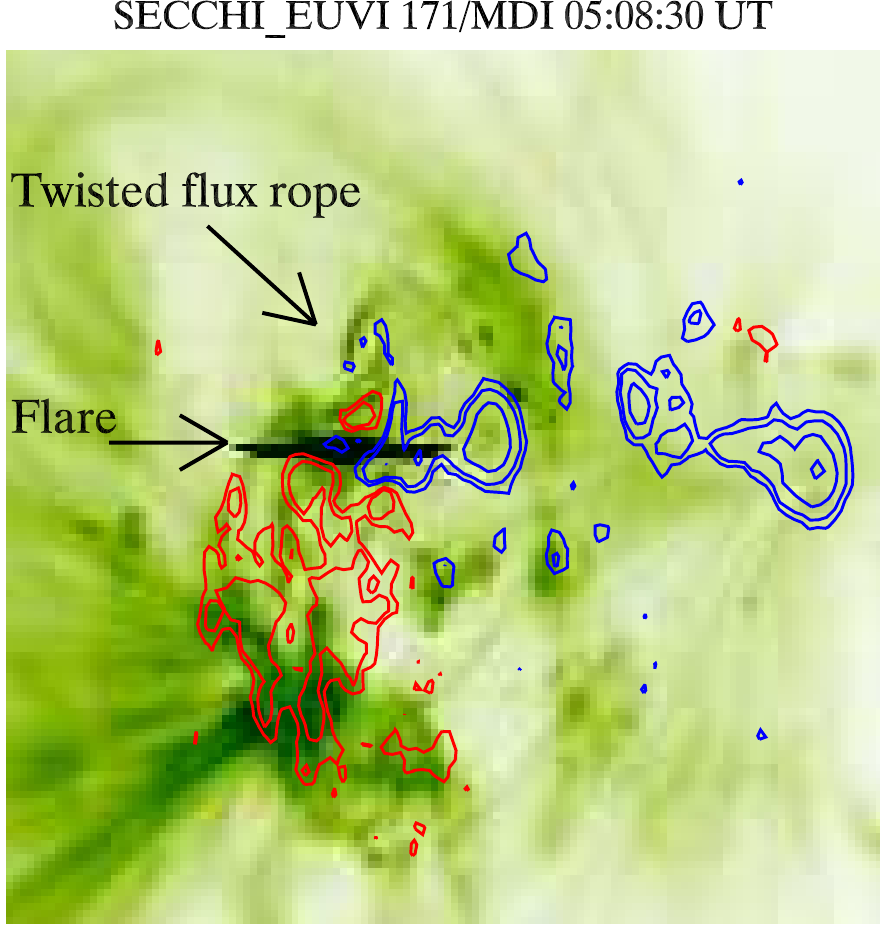}
\hspace*{-0.02\textwidth}
\includegraphics[width=0.333\textwidth]{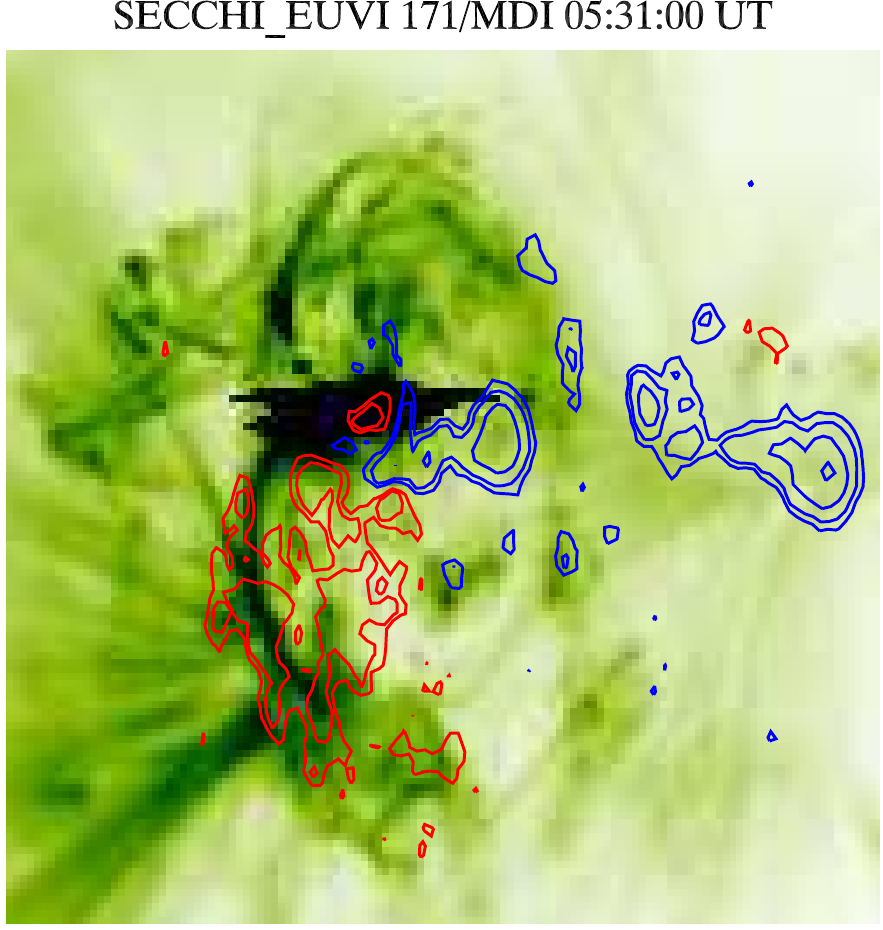}
\hspace*{-0.02\textwidth}
}
\caption{MDI contours overlaid on STEREO/SECCHI 171 \AA \ EUV images before the flare initiation and during flare progressive phase. Red contours show the positive polarity, while blue ones show the negative polarity.  The size of each image is 200$^{\prime\prime}$$\times$200$^{\prime\prime}$.}
\label{fig6}
\end{figure}
Figure 4 displays the selected SECCHI/EUVI coronal images during \,05:03--05:56\, UT. The image at 05:08 UT shows the secondary activation of helical twisted structure above the same positive-polarity sunspot which connects neighbouring, small, negative-polarity sunspot (see the middle panel of Figure 6). It should be noted that the successive activation of the helical twists, at the same place that cause the plasma brightening also, may be a signature of the build-up of magnetic energy in AR 10960. However, the activation of the secondary helical-twist at 05:08 UT seems to play crucial role in the energy build-up process for the M8.9/3B class flare. This activation is located near the centre of the active region where the flare initiation takes place, as it spreads away from the sunspot. The flare reaches maximum at $\approx$05:14 UT and then decays slowly. During the decay phase of the flare, the image at 05:23:30 UT shows a twisted flux rope (one turn is visible) very close to the flare energy-release site (indicated by an arrow). This structure moves away very slowly and finally disappears at 05:48 UT. However during the maximum and decay phase of the flare, several twisted structures are visible which indicate the presence of free magnetic energy in the observed coronal volume. We overlaid the MDI positive (red) and negative (blue) polarity contours over the selected SECCHI images to view the field morphology, during the flare event (see Figure 6). The careful investigation of these images reveals the association of  secondary helical twisted structure at 05:08 UT before the M-class flare maximum, associated with the small positive-polarity satellite sunspot. Activation of this twist is also clearly visible in SOT images. This structure seems to be moving to the north of the sunspot. The twist angle can be estimated by measuring the number of turns in the twisted helical structure. One turn corresponds to the twist angle of 2.0$\pi$. It is evident from STEREO/SECCHI images (see Figure 5) that the secondary helical twisted structure at 05:08 UT show minimum of $\approx$2.0 turns, suggesting the total twist angle probably crosses the critical limit of 2.5$\pi$ ({\it i.e.} the total twist angle is 4$\pi$). The flare brightening takes place in between a small positive and surrounding negative-polarity regions.

The rather small scale of the twisted coronal structures does not allow us to see differences in SECCHI/EUVI images of STEREO-A and STEREO-B spacecrafts with separation of only near 10$^\circ$ in June 2003 as it is able to do for more large scale erupting structures (\opencite{gis2008}; \opencite{lie2009}). Figure 5 shows enlarged images of the twisted structure obtained by both STEREO A and B spacecraft. The upper parts of the helix are aligned with visible coronal loopdf, while the bottom parts are assumed to keep the same pitch angle. According to this assumption, some two turns of the helix can be recognized especially within the secondary twisted structure (bottom panel of Figure 5). In fact it looks like a multiple-thread screw and it is difficult to follow one particular thread through the whole flux rope length. Approximate conservation of the pitch angle can help to estimate the amount of the total twist.  The primary helical twist occured during \,04:42--04:51\, UT, given in the upper panel for comparison with the secondary helical-twisted structure. The primary helical twist presented here for comparison is found to be the triggering mechanism of the B5.0 class flare during \,04:40--04:51\, UT \cite{sri2010}.

\begin{figure} 
\centerline{
\hspace*{-0.02\textwidth}
\includegraphics[width=0.33\textwidth]{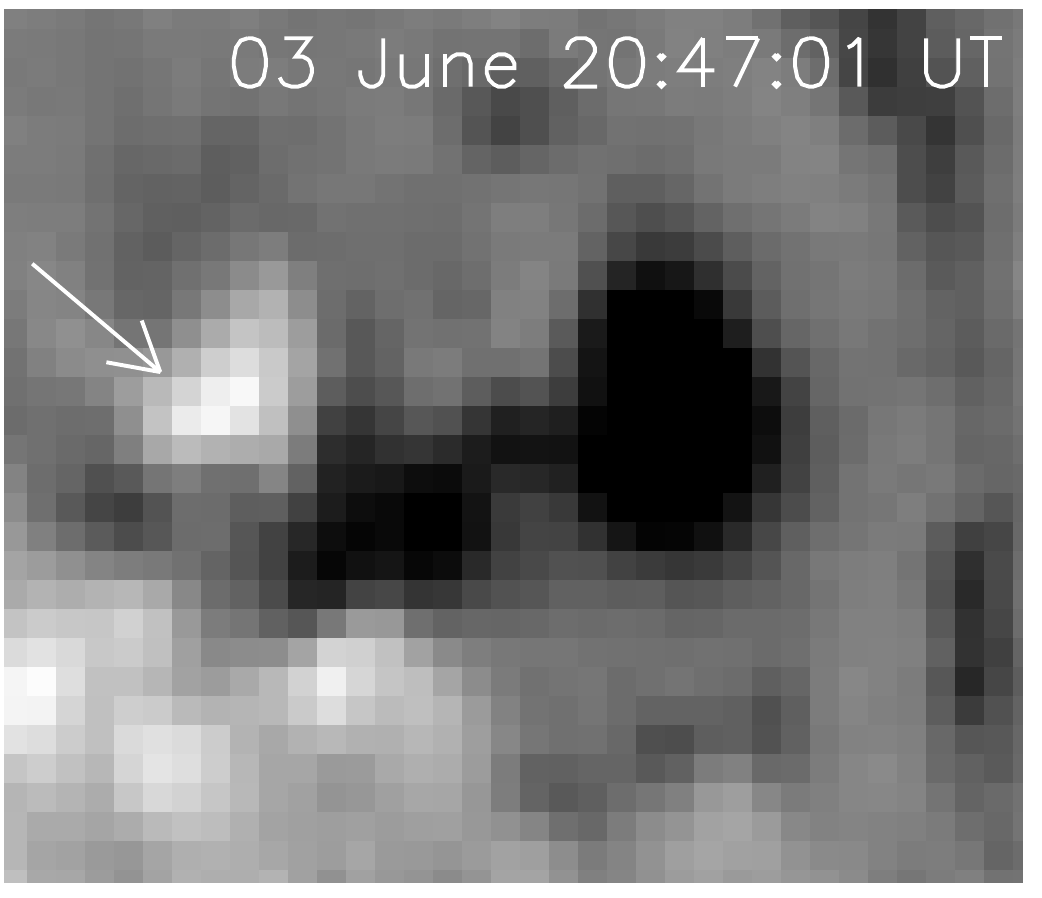}
\hspace*{-0.02\textwidth}
\includegraphics[width=0.34\textwidth]{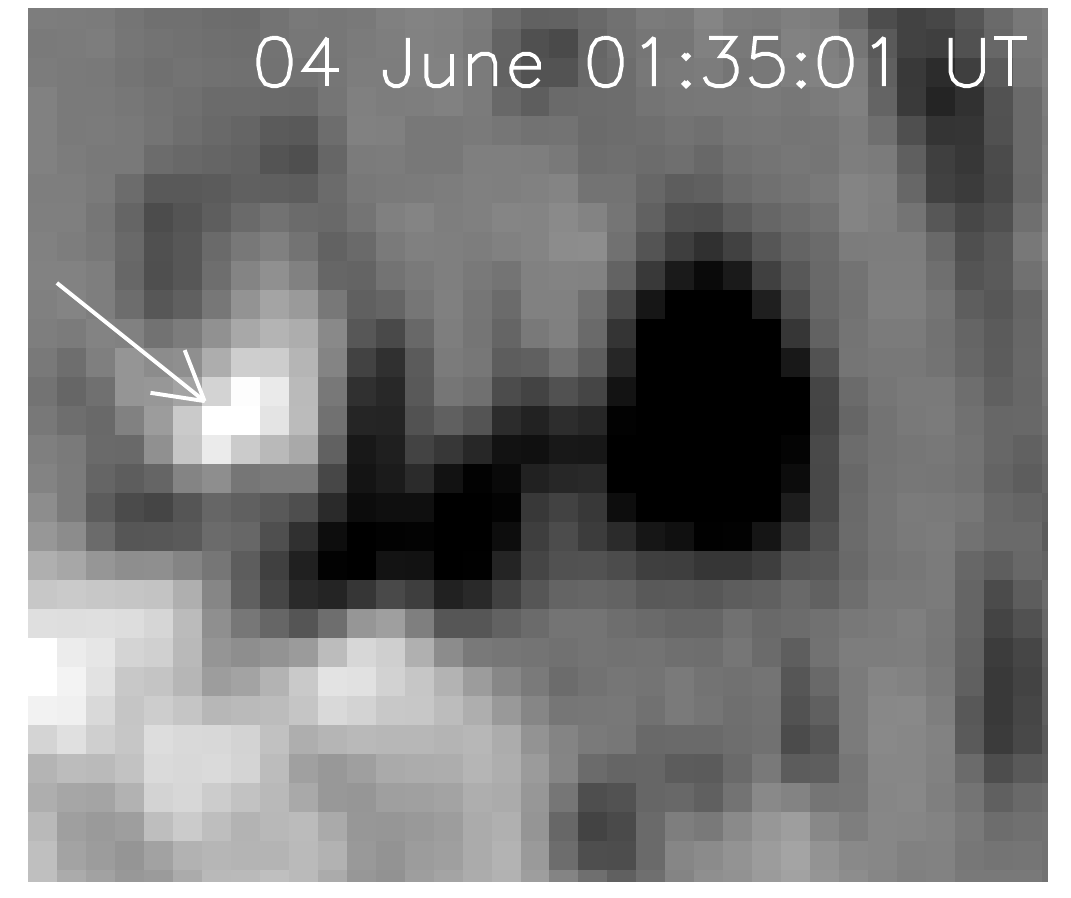}
\hspace*{-0.02\textwidth}
\includegraphics[width=0.34\textwidth]{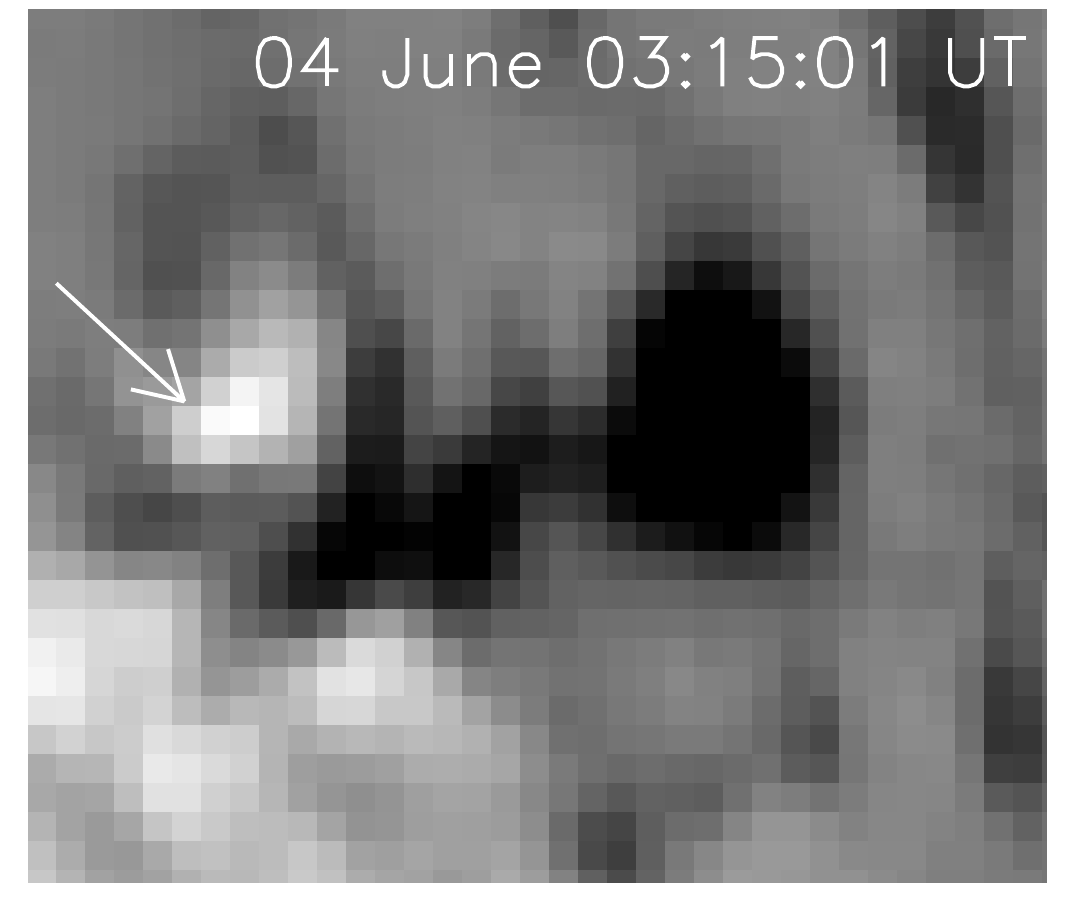}
\hspace*{-0.02\textwidth}
}
\centerline{
\hspace*{-0.02\textwidth}
\includegraphics[width=0.33\textwidth]{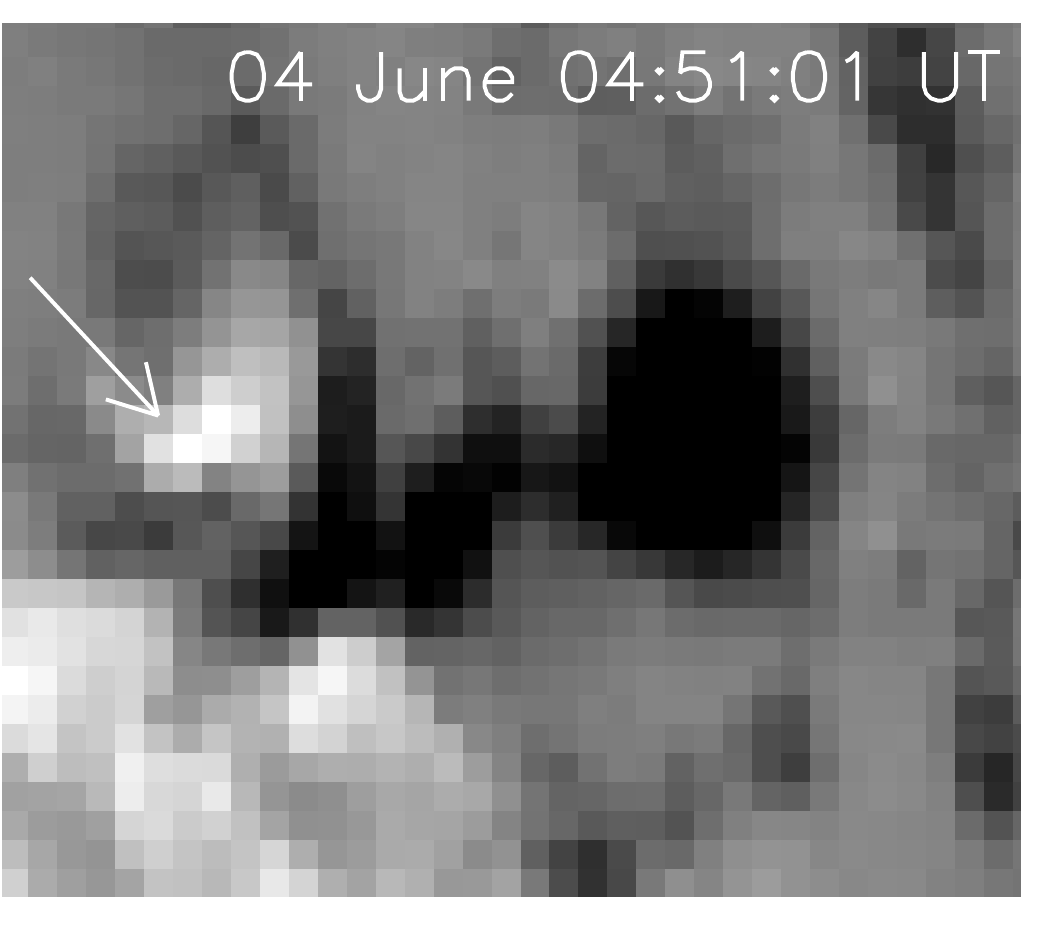}
\hspace*{-0.02\textwidth}
\includegraphics[width=0.347\textwidth]{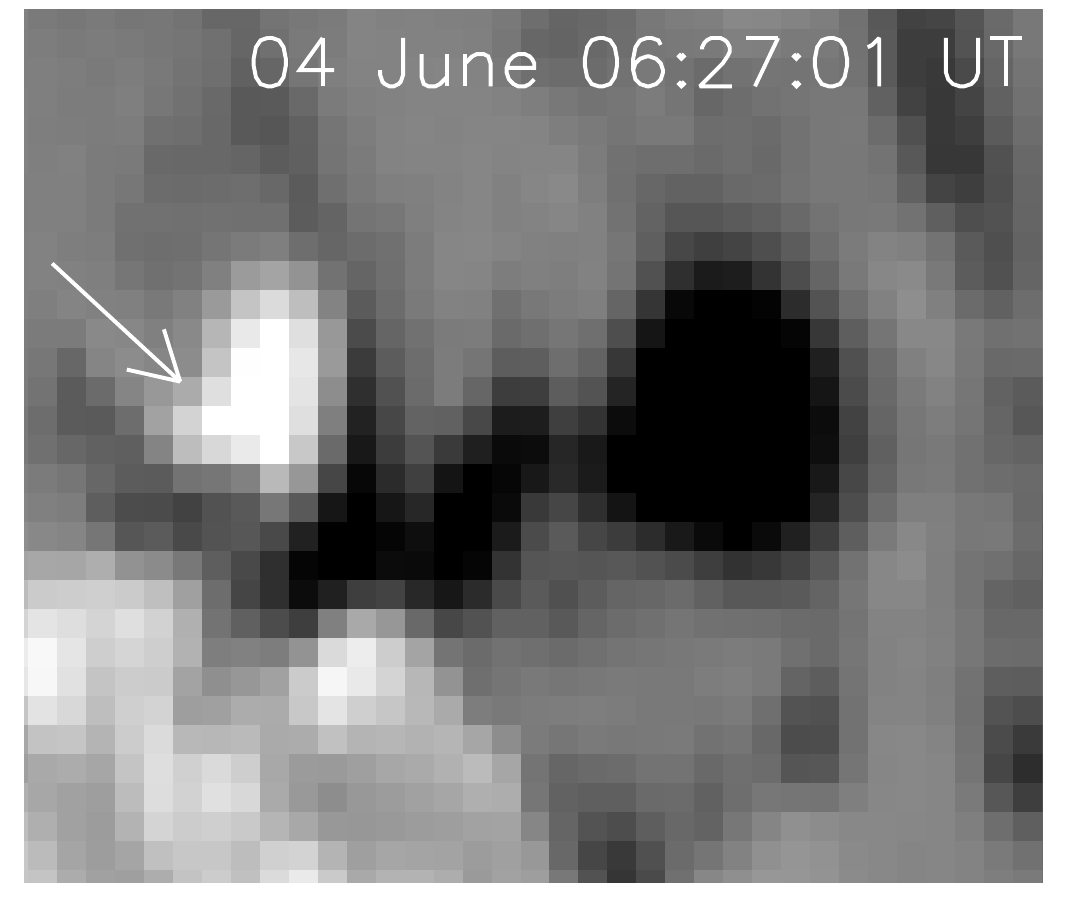}
\hspace*{-0.02\textwidth}
\includegraphics[width=0.34\textwidth]{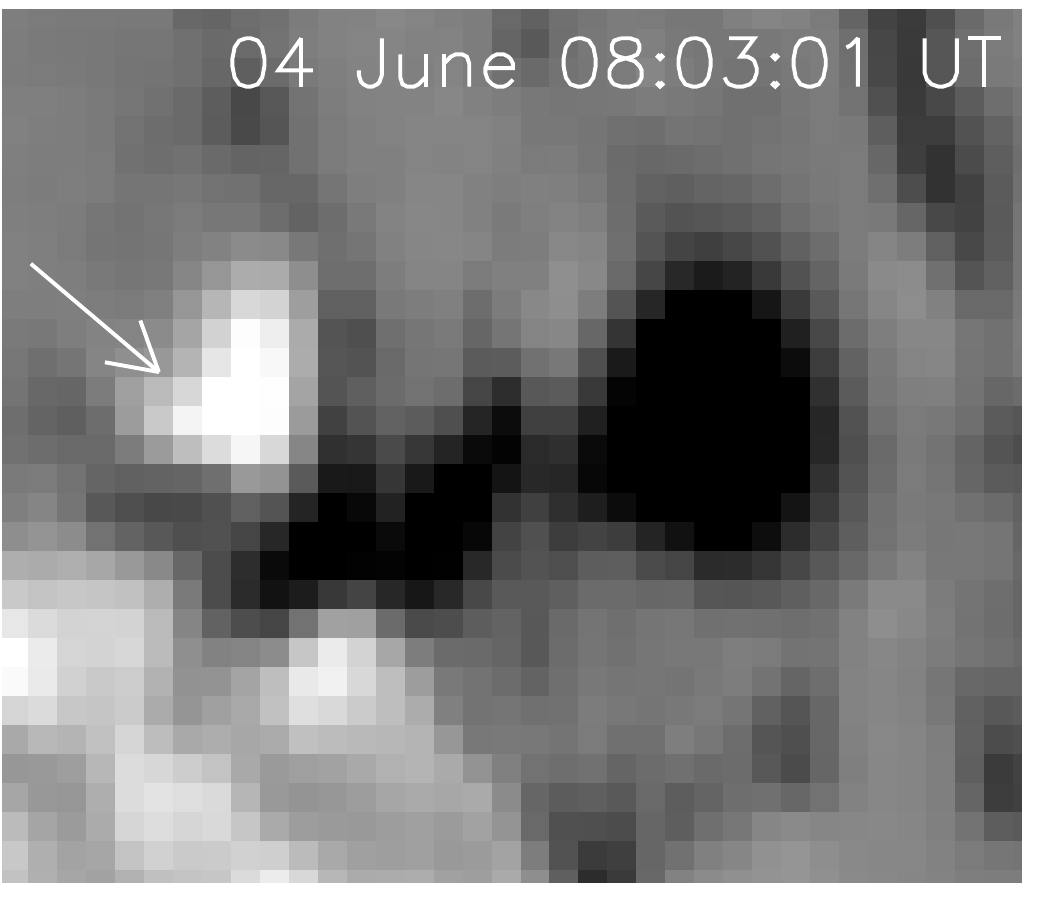}
\hspace*{-0.02\textwidth}
}
\caption{The selected SOHO/MDI images of the flare site. The arrow indicates the evolution of positive magnetic-flux region before and after the flare activity. The size of each image is 70$^{\prime\prime}$$\times$60$^{\prime\prime}$.}
\label{fig7}
\end{figure}


\section{Sunspot Evolution in SOHO/MDI and SOT/G Band Images}

We use SOHO/MDI observations to see the magnetic-field evolution during the flare.  The size of each image is 1024$\times$1024 (2$^{\prime\prime}$ per pixel resolution) with a cadence of 96 minute \cite{sch1995}. We use the standard SolarSoft library to correct the differential rotation and analyze the magnetograms.  Figure 7 displays the sequence of MDI images on 03 and 04 June 2007. The MDI movie and time-sequence images reveal the interesting features, which show the considerable changes (area enhancement) in the positive-polarity sunspot during the decay phase of M-class flare event.
\begin{figure} 
\centerline{
\hspace*{-0.02\textwidth}
\includegraphics[width=0.33\textwidth]{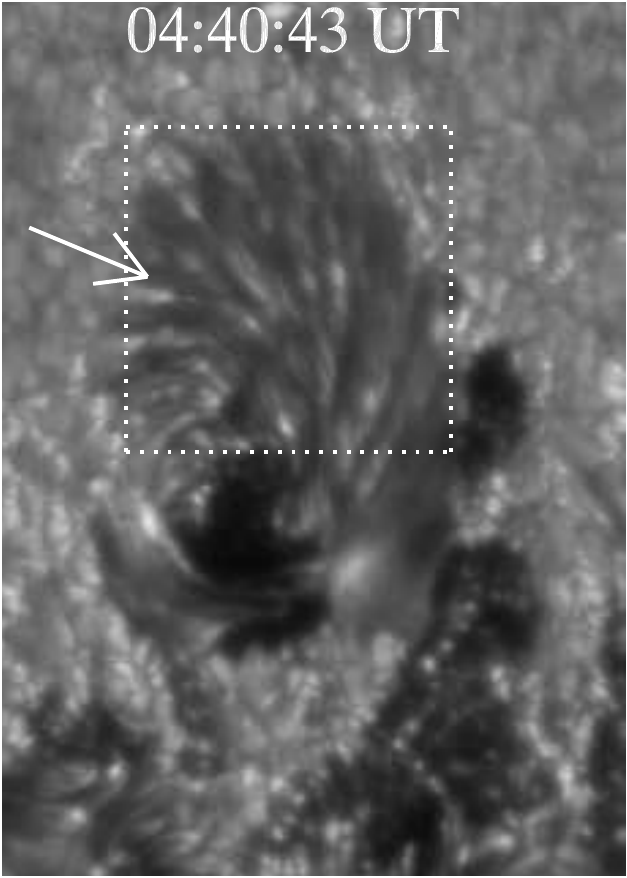}
\hspace*{-0.02\textwidth}
\includegraphics[width=0.33\textwidth]{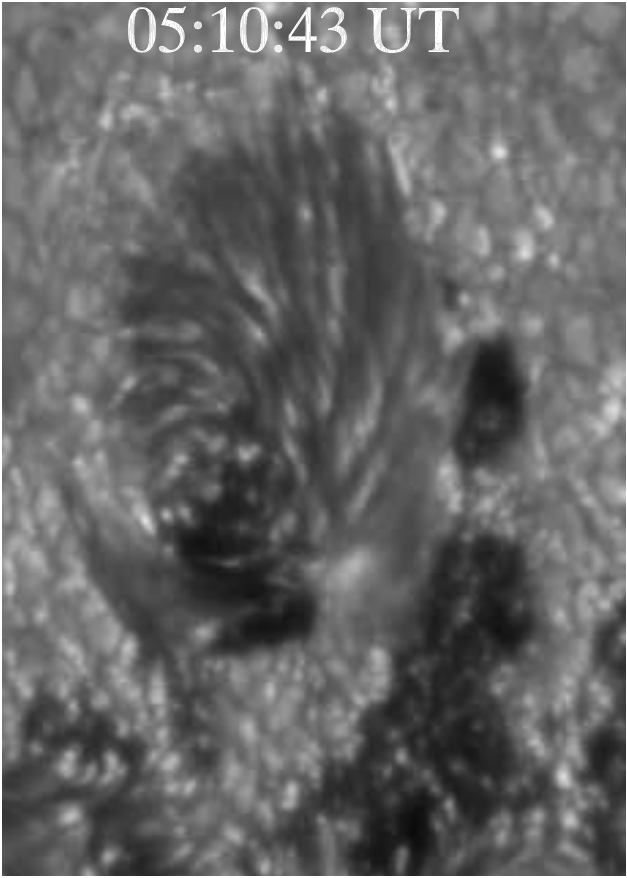}
\hspace*{-0.02\textwidth}
\includegraphics[width=0.33\textwidth]{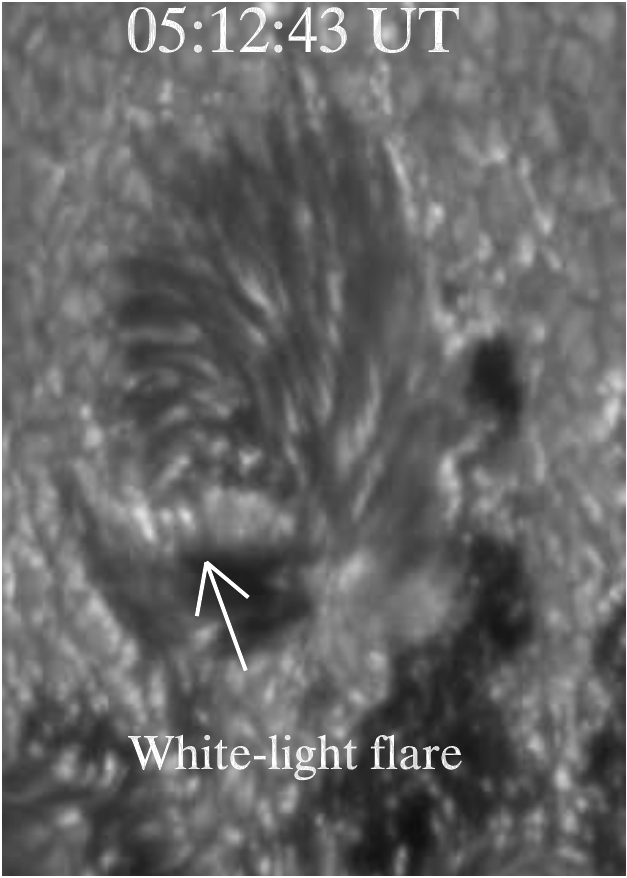}
\hspace*{-0.02\textwidth}
}
\centerline{
\hspace*{-0.02\textwidth}
\includegraphics[width=0.33\textwidth]{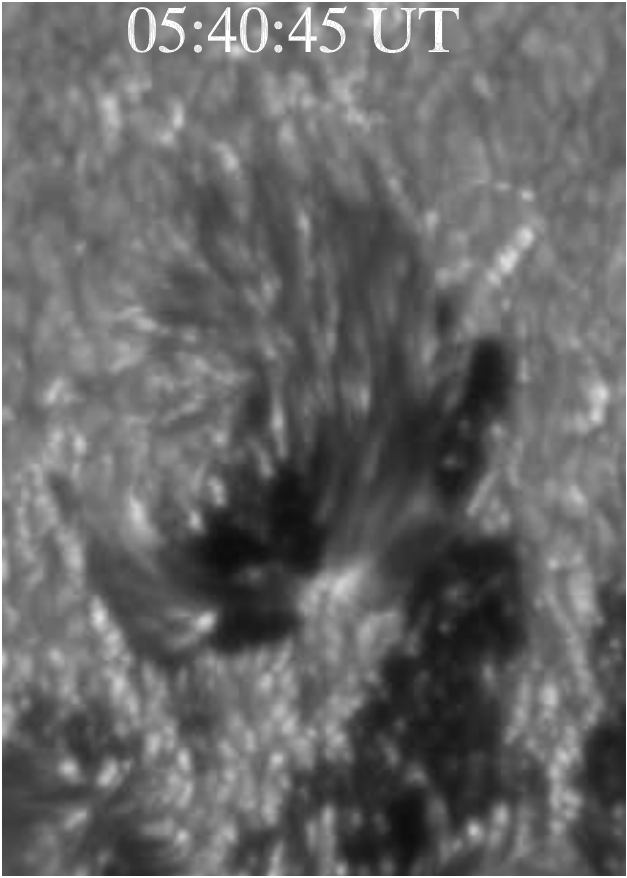}
\hspace*{-0.02\textwidth}
\includegraphics[width=0.33\textwidth]{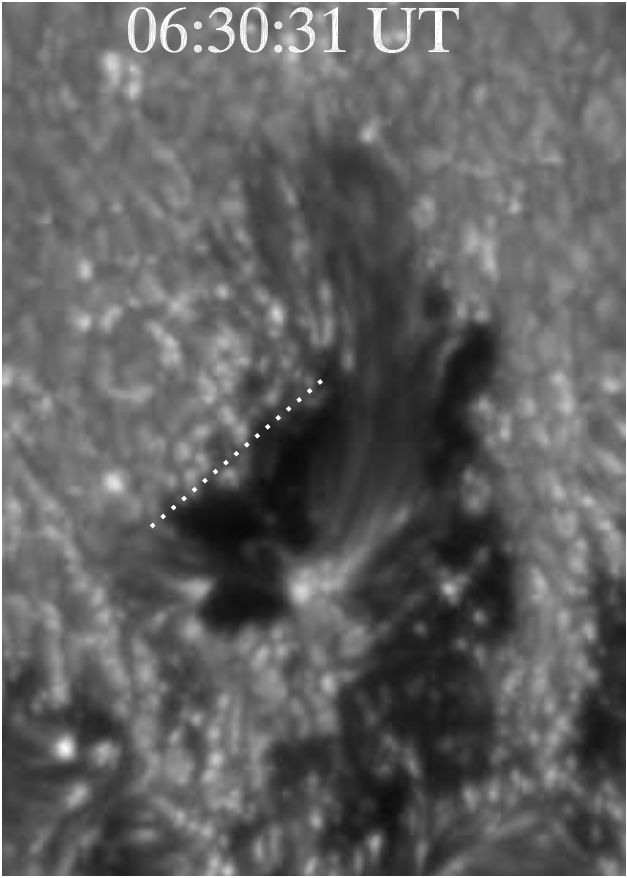}
\hspace*{-0.02\textwidth}
\includegraphics[width=0.33\textwidth]{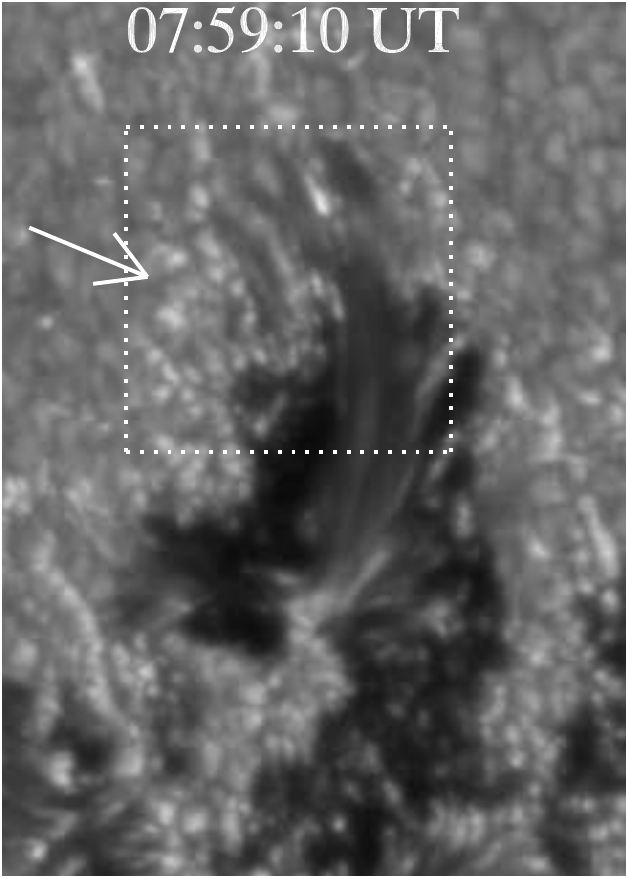}
\hspace*{-0.02\textwidth}
}
\caption{The selected SOT/G-band images (4305 \AA) showing the evolution of the positive-polarity sunspot (before, during and after the M8.9/3B class flare). The dotted line and boxes reveal the orientation change and disappearance of twisted penumbral filaments respectively after the flare event. The size of each image is 25$^{\prime\prime}$$\times$35$^{\prime\prime}$.}
\label{fig8}
\end{figure}
\begin{figure} 
\centerline{
\includegraphics[width=0.8\textwidth]{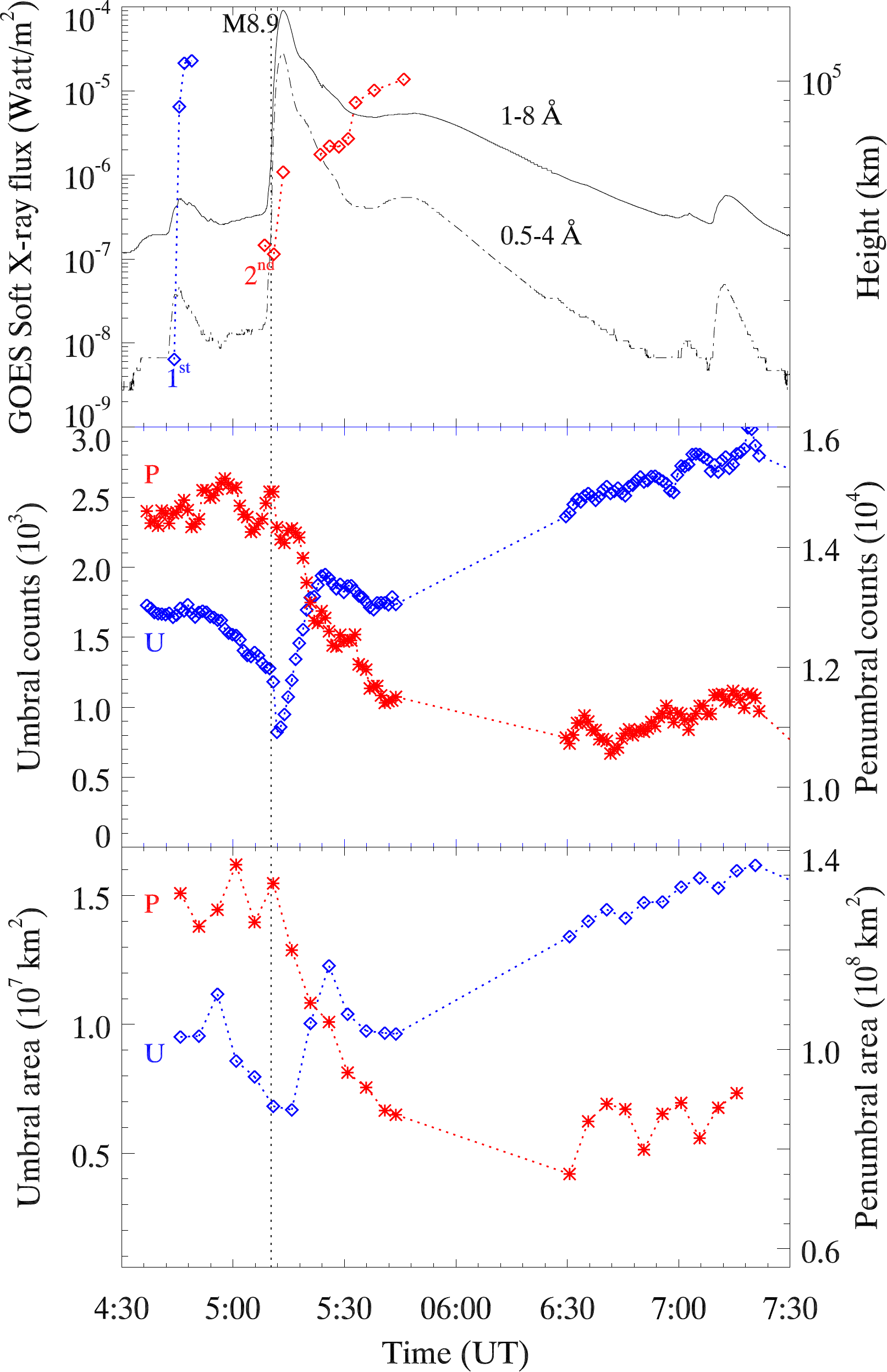}
}

\caption{Top: Projected height (elongation) {\it vs.} time profiles of both twisting helical magnetic structures with soft X-ray flux profiles of the flares on 4 June 2007. This plot clearly indicates that the rise of magnetic structures is closely associated with the flare onset. Middle and bottom: Umbral and penumbral changes (indicated by ``U'' and ``P'' respectively) in intensity and area to show the link with soft X-ray flux profiles. It is evident from the plot that there are remarkable changes (umbral enhancement and penumbral decay) in both umbra and penumbra after the flare maximum.}
\label{fig9}
\end{figure}
Figure 8 displays the selected SOT/G-band images which show the partial field of view of the active region containing the same positive-polarity sunspot as marked by arrow. This sunspot also shown by the red contour in Figure 6, which is associated with successive activation of the helical twists. The evolution of the small positive-polarity sunspot initially shows the highly twisted penumbral filaments in the counterclockwise direction ({\it i.e.} sunspot rotation is clockwise) well before the flare activity (indicated by arrow) as the active region lies in the southern hemisphere. This secondary twist at the footpoint of the loop system associated with this sunspot may be responsible for the energy build-up process of M-class flare occured in this active region. The brightening (two bright points) takes place at the opposite edges of the umbral part of this sunspot. Then, a secondary twisted helical structure rises up from the same site (refer to Section 2.1). The image observed at 05:12 UT shows the white-light flare at the sunspot (indicated by arrow) during its impulsive phase. During the decay phase of the M-class flare, we notice several changes in the sunspot: {\it i)} disappearance of twisted penumbral filaments at the northern part of the sunspot (indicated by boxes). {\it ii)} orientation change in the sunspot (shown by dotted line). Before the initiation of the flare activity, sunspot shows spherical shape with counter-clockwise penumbral filaments, whereas after the flare activity it shows the elongated shape with penumbral changes (i.e decay of penumbral filaments). 
We have estimated the projected height (elongation) {\it vs.} time profile of the twisted magnetic structures observed in TRACE and SECCHI measurements, following the apex of the structure from the centre of the active region and plotted against soft X-ray flux measurements (Figure 9, top). It is evident from the plot that the flare is closely associated with the activation and rising motion of the twisted magnetic structures/flux ropes. The co-temporal enhancement of the GOES soft X-ray flux profile with the increase of the projected height of the primary helical structure validate the findings of \inlinecite{sri2010} that this twist was probably responsible for B5.0 class flare. While again the co-temporal enhancement of the soft X-ray flux profile with the increase of the height of the secondary, helical twisted structure shows its association with the M8.9/3B flare event that occured in AR 10960.
  We have used the SOT/blue continuum (4504 \AA) images for quantitative estimation of penumbral and umbral changes during the flare event. We selected a box of 16$^{\prime\prime}$$\times$22$^{\prime\prime}$ covering the sunspot of delta configuration. We extracted the total counts less than 700 for umbral change and between \,700--1400\, for penumbral changes. For viewing the change in area of both umbra and penumra, we draw the umbral and penumral boundaries using standard routines in IDL libraries and then extracted the total number of umbral and penumbral pixels of the sunspot. Figure 9 displays the temporal changes in umbral and penumbral intensity and area with respect to the soft X-ray flux profile. It is evident from the figure that there is a remarkable changes in umbral and penumbral structures. However, the rapid change in umbral portion at $\approx$05:12 UT is due to the flare that covers the sunspot umbral part. After the flare maximum, we observe considerable penumbral disappearance ($\approx$35\,--\,40\%\,) and enhancement ($\approx$45\,--\,50\%) in the umbral area. This enhancement of the positive-polarity spot is evident in the SOHO/MDI magnetograms during the decay phase of the M-class flare (see Figure 7). This suggests that the magnetic field becomes more vertical from the initial horizontal configuration, which is in agreement with the previous studies \cite{wang2004,liu2005}. These changes suggest the loss of magnetic energy with sunspot evolution and that energy seems to be released in the form of flare thermal energy.
 

\begin{figure} 
\centerline{
\includegraphics[width=0.8\textwidth]{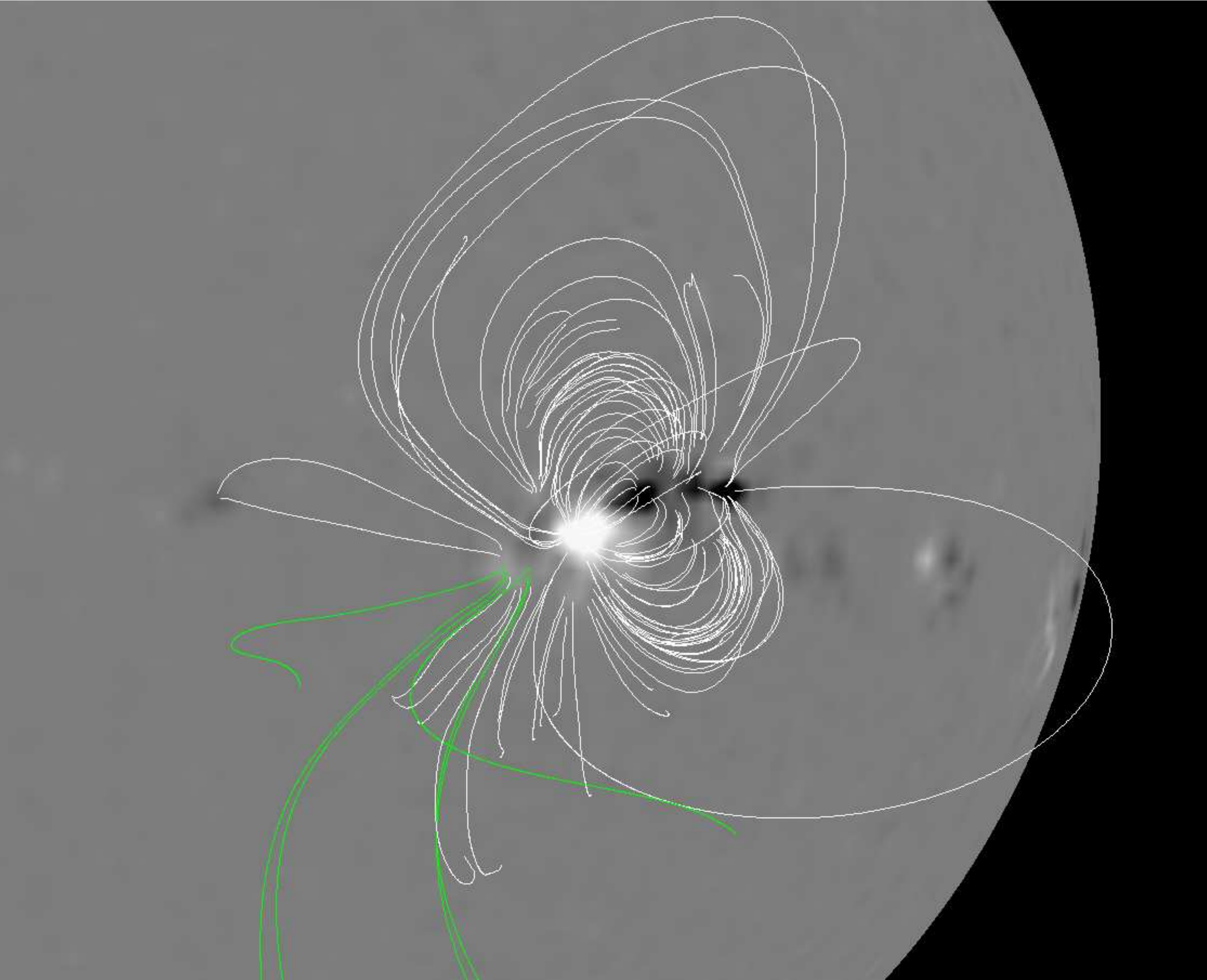}
}

\caption{Potential field source surface (PFSS) extrapolation of the NOAA AR 10960 at 00:04 UT on 4 June 2007. White lines show the closed magnetic fields whereas green lines show the open fields.}
\label{fig10}
\end{figure}
\begin{figure} 
\centerline{
\hspace*{-0.02\textwidth}
\includegraphics[width=0.5\textwidth]{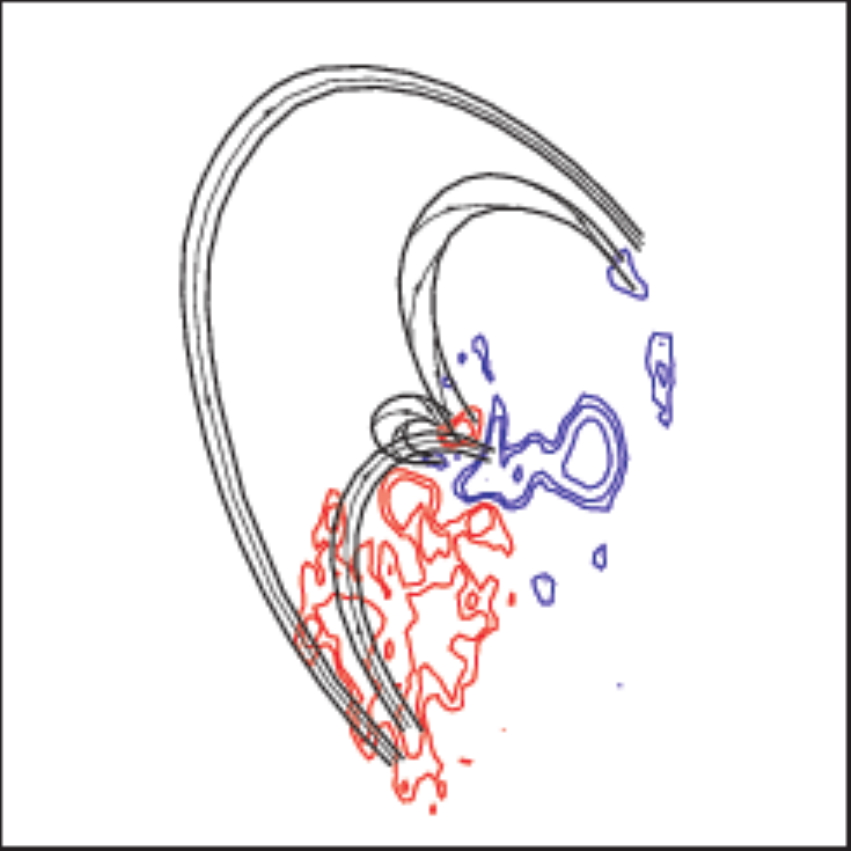}
\hspace*{-0.02\textwidth}
\includegraphics[width=0.5\textwidth]{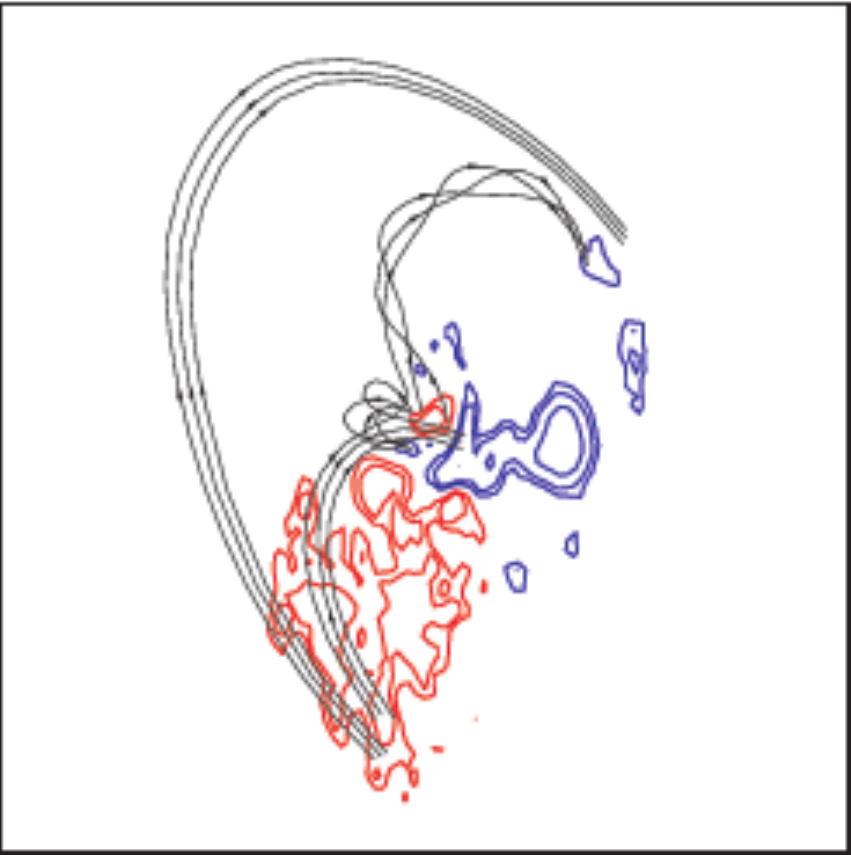}
\hspace*{-0.02\textwidth}
}
 \vspace{-0.5\textwidth}   
     \centerline{\Large \bf     
      \hspace{0.0 \textwidth}  \color{black}{(a)}
      \hspace{0.415\textwidth}  \color{black}{(b)}
         \hfill}
     \vspace{0.45\textwidth}    

\centerline{
\hspace*{-0.02\textwidth}
\includegraphics[width=0.5\textwidth]{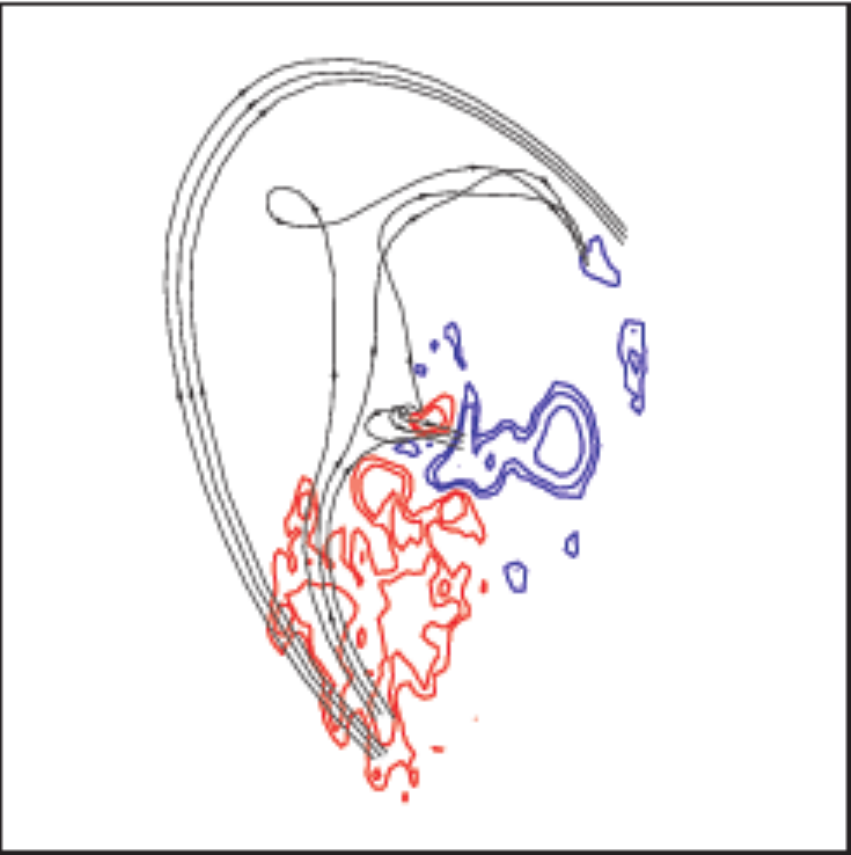}
\hspace*{-0.02\textwidth}
\includegraphics[width=0.5\textwidth]{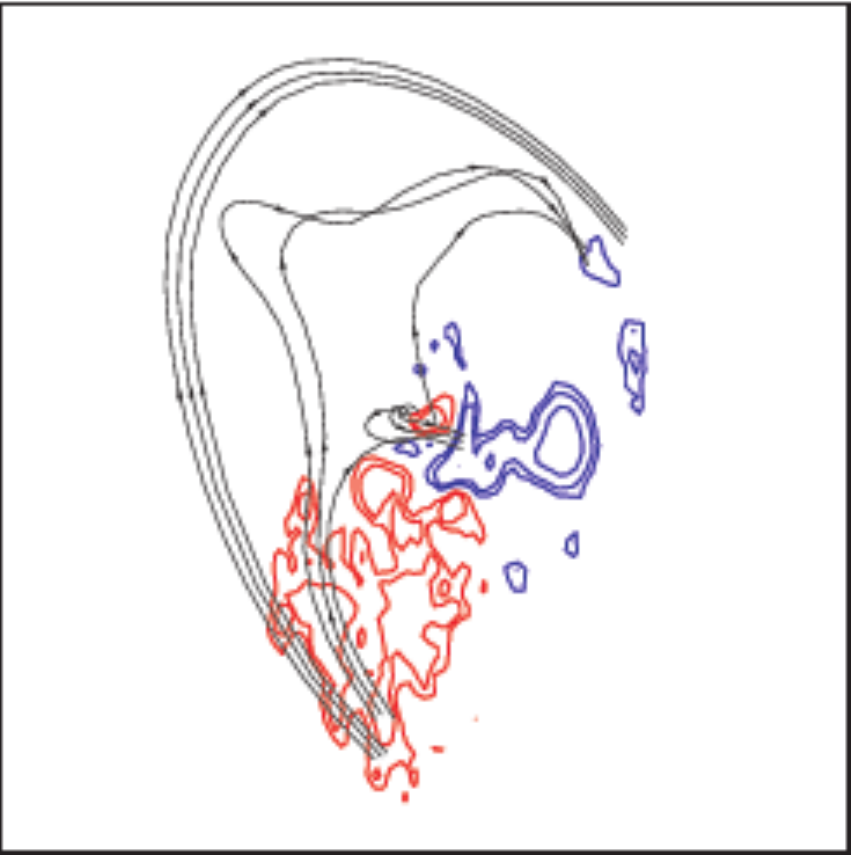}
\hspace*{-0.02\textwidth}
}
 \vspace{-0.5\textwidth}   
     \centerline{\Large \bf     
      \hspace{0.0 \textwidth} \color{black}{(c)}
      \hspace{0.415\textwidth}  \color{black}{(d)}
         \hfill}
     \vspace{0.45\textwidth}    

\caption{Schematic cartoons demonstrating the magnetic configuration of the active region before and during the flare event. Red contours show the positive-polarity sunspots whereas blue ones indicate the negative-polarity sunspots.}
\label{fig9}
\end{figure}

\begin{figure} 
\centerline{
\includegraphics[width=0.8\textwidth]{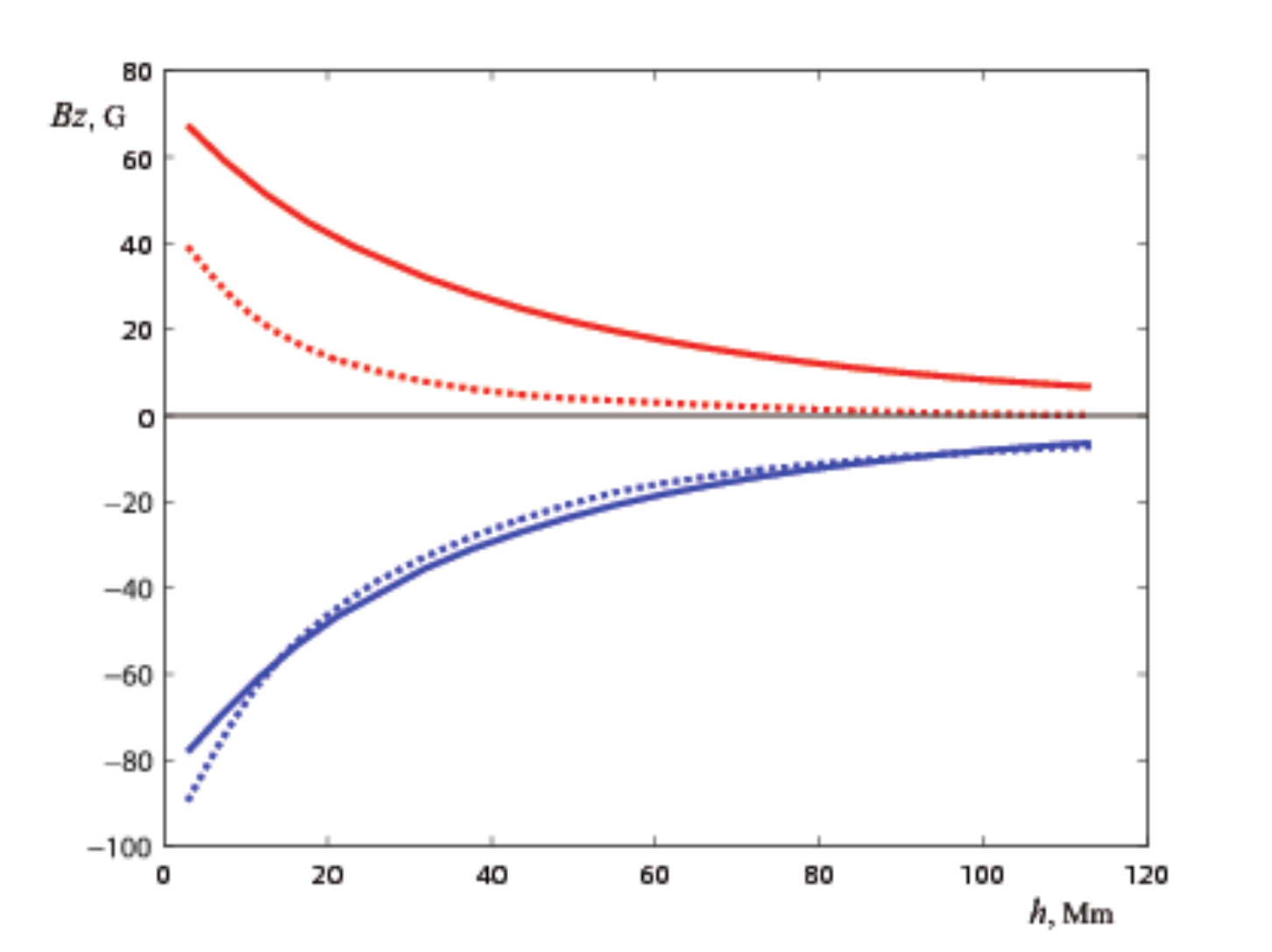}
}

\caption{Dependence of average vertical magnetic field at different heights for the active regions NOAA 10501 (dotted curves) and NOAA 10960 (thick curves). Red and blue curves
correspond respectively to the positive  and negative magnetic field strength.
}
\label{fig10}
\end{figure}
\section{Discussion}
We study the M8.9/3B flare event on 4 June 2007 from NOAA AR 10960 using multiwavelength observations. It is shown that the small positive-polarity sunspot plays an important role in triggering this flare event, which lies at the centre of the active region and it is associated with the twisted flux tube/rope where successive helical twists have been activated before the flare event. MDI observations also reveal the considerable amount of enhancement in the area of the positive-polarity sunspot at the flare site during the decay phase of the M-class flare which is in agreement with the SOT/G-band observations. Another interesting point is that the positive-polarity sunspot seems to be  highly sheared ({it i.e.} twisted penumbral filaments) before the flare activity.  During the decay phase and after the flare event, the sunspot indicates major changes in its structure. It shows twisted penumbral filament disappearance in the northern part of the sunspot. Before the flare event, the penumral structure was highly sheared showing anticlockwise orientation. After the flare event, it becomes more simplified with the loss of some area. This suggests the loss of twisted magnetic energy associated with the sunspot, which may be released in the form of flare thermal energy.
 For investigating the overlying magnetic field environment of this active region, we have used the potential-field source surface (PFSS) extrapolation \cite{alt1969,sch1969} before the flare event at 00:04 UT (see Figure 10). We rotate the SOHO/MDI image for our convenience to see the morphology of the magnetic field. However, the active region lies near the eastern limb during that time. The coronal magnetic field topology is on average in agreement with TRACE and SECCHI observations.

Figure 11 shows a  schematic scenario of the event, which we have deduced from the multiwavelength analysis. EUV images reveal several flux tubes that seem to play a major role in the progress of the flaring activity. There are large loopdf that straddle the whole active region. They connect the southern border of the large area of positive-polarity and small fragments of negative-polarity to the North of the major sunspots. Two smaller flux tubes originate from the vicinities of the footpoints of the large loop but end at the centre of the active region. The northern flux tube connects with the positive-polarity sunspot, while the southern flux tube connects with the negative-polarity magnetic flux concentrations nearby the sunspot. There is also a short flux tube connecting the dominant, positive polarity of the sunspot with the small portion of its umbra with opposite-polarity. Highly twisted structure of the sunspot penumbra indicates the presence of the twist within the short flux tube and the northern flux tube. Although, the most clear twisting of the northern flux tube becomes visible after its activation. The character of the brightening propagation allows to make conclusion that the twist within the northern flux tube does not appear during the flare activity, but existed long time before the flare event. Therefore, the positive-polarity sunspot is the footpoint of the flux rope. Some part of this magnetic flux tube/rope is connected with distant photospheric negative elements and some part is connected with the negative portion of the same umbra. After the activation, caused possibly by the emergence of a new magnetic flux which manifests itself in our region of interest as a growth of the area of positive-polarity (Figure 7), several field lines of the flux rope may reconnect with the field lines of the southern flux tube. As a result, a new, long flux tube is created as well as short loopdf also that connect the former central footpoints of reconnecting field lines (Figure 11b). This short loop corresponds to the post-flare loop system visible in SOT/Ca {\sc ii} H line images. Reconnected, long field lines move up and the twist propagates from the flux rope along the whole length of the flux tube. The field lines form a loop firstly or a noose  (Figure 11c), and then make up a wide tangled structure (Figure 11d). It is most likely that the presence of the twist causes the field lines to move up, however they stop at some higher altitude. Therefore, the whole scenario may also resemble a failed flux-rope eruption  \cite{filippov2002,ji2003}. After the flare, the magnetic-field configuration becomes simplified. There is no longer a clear manifestation of the twisted magnetic field. Field lines of the positive-polarity sunspot become more connected to the nearby negative polarities to the right of the sunspot. This possibly results in disappearance of penumbral filaments in the northeast part of the sunspot. 

SOT and TRACE observations indicate the successive activation of helical, twisted flux bundles just above the same positive-polarity sunspot (at the edges of the twisted umbral structure) well before the flare event.  As the secondary activation of the helical twist rises at 05:08 UT, the M-class flare intensity maximizes. This can be interpreted as the rising of twisted flux rope and its progressive reconnection with the surrounding opposite-polarity field region. The plasma is heated up, evaporated, and pumped into the two smaller loopdf (underlying a major loop system), which connect to the reconnection site with opposite magnetic polarities (refer to SECCHI images during the decay phase of the flare).

\inlinecite{sri2010} have observed the first activation of a highly (right\,--\,handedely) twisted flux tube in AR 10960 during the period 04:43\,--\,04:52 UT. They have estimated the length and the radius of the loop as $L\approx$80 Mm and $a\approx$4.0 Mm
respectively, and also estimated total maximum twist angle as $\Phi\approx$12$\pi$, by assuming quasi-symmetric distribution of the twist over the magnetic loop, which is much larger than the Kruskal--Shafranov instability criterion. They have found that right-handed twist may be asymmetrically distributed over the observed loop, which is smoothed with the Alfv\'enic time $\approx$80 seconds and possess a quasi-symmetric maximum twist. The detection of clear double structure of the loop top during 04:47\,--\,04:51 UT in TRACE 171 \AA \ images, are found to be consistent with simulated kink instability in curved coronal loopdf \cite{torok2004}. They have suggested that the kink instability of this twisted magnetic loop triggered the B5.0 class solar flare, which also occurred between 04:40\,--\,04:51 UT in this active region. The co-spatial brightening in soft X-ray as observed by {\it Hinode}/XRT and the co-temporal occurrence of the right-handed twisting in the flux tube confirm the occurence of the B5.0 flare during 04:40\,--\,04:51 UT probably due to the generation of the kink instability. We have also found the secondary twist of the same handedness (right-handed) in the magnetic flux tube at the same place as in the STEREO images at 05:08 UT (refer to Figure 5). Therefore, the activation of a second helical twist may also be associated with the kink instability in the active region which may trigger the M-class flare on 05:08 UT. In Figure 5, we present the twisted flux tube observed by TRACE, STEREO A and B. We find the activation of the secondary twist on the observed loop with same right handedness and $\approx$two turns in the flux tube ({\it i.e.} $\approx$4$\pi$ twist)  at first halve of the loop (see bottom panel of Figure 5). This is also
crossing the threshold of minimum twist for the stability in the 
flux tube.

\inlinecite{ishii1998} also have reported that the flare productive magnetic shear is produced by the emergence of twisted magnetic flux bundle. Magnetic energy is stored in the twisted flux bundle, which is originally formed in the convection zone and released as flares in the course of the emergence of the twisted flux bundle above the photosphere. Our study provides the most likely signature of the successive activation of the helical twist in the flux tubes/ropes moving away from the active region. The location of the flare activity coincides with the area of the positive magnetic flux increase. Therefore, it may be concluded that the magnetic energy stored in the helical-twisted flux bundles is released as a flare by reconnecting with the surrounding opposite-polarity field lines, when it moves away from AR. 
This active region produces M-class flare without any coronal mass ejections (CMEs) on 4 June 2007. According to analytical and numerical models of magnetic flux rope eruption, the behavior of the flux rope strongly depends on the rate of ambient magnetic field decrease with height \cite{van1978,forbes1995,forbes1995,torok2004,torok2005}. If the magnetic field is strong enough at high altitude and has a significant horizontal component, the ascending motion of a flux rope can be stopped at a greater height and does not lead to formation of a CME. We compared the coronal potential magnetic field of the active region NOAA 10960 with the magnetic field of active region NOAA 10501 studied by \inlinecite{kumar2010}, which produced a fast full-halo CME. Figure 12 shows the averaged vertical magnetic field [B$_z$] as a function of height [h] for both active regions. The area of about 220$\times$240 Mm from MDI magnetograms was used as a boundary condition for numerical solving the Neumann external boundary-value problem \cite{sch1964,filip2001}. Magnetograms on 19 November 2003 and 7 June 2007 were chosen for the dates when the active regions were close to the center of the solar disk. The area 220$\times$240 Mm covers all marked photospheric magnetic fields of the active regions. Positive and negative magnetic flux was calculated at different heights and then divided by the area occupied by this flux to obtain magnetic flux density or averaged magnetic-field strength. The dependence of average vertical magnetic field at different heights for the active
regions NOAA 10501 (dotted curves) and NOAA 10960 (thick curves) is compared in this Figure 12 in which
red and blue curves respectively correspond to the positive and negative magnetic field strength.

It is evident that magnetic field in the CME-productive AR falls more rapidly with height than in the other one and the magnetic flux is unbalanced. Positive flux is negligible above 30 Mm that is evidence of weak horizontal field needed to retard the flux rope ascending motion. In contrast, positive and negative fluxes in NOAA 10960 are nearly equal and decrease slowly and synchronously with height. This means the presence of large-scale, closed magnetic field that is able of supporting the flux-rope equilibrium at high altitude. As it was stressed by \inlinecite{torok2005} the decrease of the overlying field with height is a main factor in deciding whether kink instability leads to a confined event or to a CME.

Therefore, the large-scale destabilization of AR 10960 magnetic fields does not occur during this flare. Our multiwavelength and high-resolution observational results from recent spaceborne instruments reveal the dynamics of this active region and the associated M-class flare, which may be unique evidence for further theoretical modeling and observational studies of those active regions that produce solar flares but are the poor originators of CMEs. These observational evidences may also be further useful in the forcasting of the occurrence of large-scale solar eruptive phenomena from similar kinds of active regions. However, further statistical, multiwavelength studies should be carried out using forthcoming high-resolution space and ground-based observations to investigate the dynamics and mechanism respectively for multiple active regions and associated flares, which do not produce any large-scale CMEs. These studies may be useful to find the detailed physical scenario and dynamics of such unique active regions.

\section{Conclusions}
We find multiwavelength evidence of the successive activation of helical twists that may help in the energy build-up process at the flare site. The built energy is released later in the form of M-class flare after secondary activation of this critical twist in the flux tube/rope and its reconnection with neighbouring opposite fields. The main conclusions of this study may be summarized as:

{\it i)}. We report the dynamics of a single positive-polarity sunspot having twisted penumbral filament structure and successive activation of twisted helical magnetic structures.

{\it ii)}. The activation of two helical structures/ropes played an important role in destabilizing the field lines and in triggering the flare. The twist in the secondary magnetic structures crosses the threshold limit (2.5\,--\,3.5$\pi$), which probably produces the kink instability in this structure. The energy-release site of the M-class flare {\it i.e.} at the center of the AR, coincides with the activated twisted magnetic structures. \inlinecite{ishii1998} pointed out that the successive emergence of helical flux bundles plays crucial role in triggering flares and showed it by a schematic cartoon, but here we provide the observational evidence of the same.

{\it iii)}. The M-class flare shows agreement with the quadrupolar (closed-closed) reconnection model (breakout) between two closed field lines \cite{anti1998}. The asymmetric evolution is driven by footpoint shearing of one side arcade, where reconnection between the sheared arcade and the neighboring (unsheared) flux system triggers the flare. As the twisted magnetic structure moves away from the reconnection site, the flare intensity increases and reaches to maximum. This reveals the progressive reconnection of the twisted magnetic structure with the surrounding opposite-polarity fields.

{\it iv)}. Penumbral disappearance during the decay phase and after of the flare event suggests that the magnetic-field structure to becomes more vertical. This indicates that the magnetic field changes from a highly inclined to an almost vertical configuration during the decay phase of the flare (just after the flare maximum), {\it i.e.} part of penumbral magnetic field is converted into umbral fields. Our results are in agreement with previous studies \cite{wang2004,liu2005}.

{\it v)}. The decrease of the overlying field with height is a main factor in deciding whether the kink instability leads to a confined event or to a CME \cite{torok2005}. In AR 10960, the slow variation of vertical component of the magnetic field with height is found to be the most likely the cause for the failed eruption during M-class flare.

\begin{acks}
We thank the reviewers for their valuable suggestions which improved the manuscript considerably. We acknowledge SOHO/MDI and TRACE for providing the data used in this study.  SOHO is a project of international cooperation between ESA and NASA. We acknowledge the Hinode and STEREO missions for providing us the high-resolution data. 
 Hinode is a Japanese mission developed and launched by ISAS/JAXA, with NAOJ as
 domestic partner and NASA and STFC (UK) as international partners. It is operated
 by these agencies in co-operation with ESA and NSC (Norway). This work was supported in part by the Department of Science and Technology, Ministry of Science and Technology of India and in part by the Russian foundation for Basic Research (grants 09-02-00080 and 09-02-92626, INT/RFBR/P-38).  
\end{acks}

\bibliographystyle{spr-mp-sola}
\bibliography{references}  
\end{article}
\end{document}